\documentclass{emulateapj}

\newcommand{\apm}{APM\,08279+5255}
\usepackage{amssymb}

\def\ts     {\thinspace}
\def\kms    {\ts km\ts s$^{-1}$}
\def\etal   {{\rm et\ts al.}}

\def\ci     {C\ts {\scriptsize I}}
\def\civ    {C\ts {\scriptsize IV}}
\def\nv     {N\ts {\scriptsize V}}

\def\aco    {{\rm CO}($J$=1$\to$0)}
\def\bco    {{\rm CO}($J$=2$\to$1)}

\def\dco    {{\rm CO}($J$=4$\to$3)}

\def\fco    {{\rm CO}($J$=6$\to$5)}

\def\ico    {{\rm CO}($J$=9$\to$8)}
\def\jco    {{\rm CO}($J$=10$\to$9)}
\def\kco    {{\rm CO}($J$=11$\to$10)}

\def\ehnc    {{\rm HNC}($J$=5$\to$4)}

\def\acn    {{\rm CN}($N$=1$\to$0)}
\def\dcn    {{\rm CN}($N$=4$\to$3)}

\slugcomment{draft version \today, accepted for publication in the Astrophysical Journal}

\shorttitle{High-Resolution CO Imaging of APM\,08279+5255}
\shortauthors{Riechers et al.}

\begin{document}

\title{Imaging the Molecular Gas in a $z$=3.9 Quasar Host Galaxy at
  0.3$''$ Resolution: \\
  A Central, sub-kiloparsec Scale Star Formation Reservoir in
  APM\,08279+5255}

\author{Dominik A. Riechers\altaffilmark{1,2,5}, 
  Fabian Walter\altaffilmark{1}, Christopher L. Carilli\altaffilmark{3}, and
  Geraint F. Lewis\altaffilmark{4}}

\altaffiltext{1}{Max-Planck-Institut f\"ur Astronomie, K\"onigstuhl 17, 
Heidelberg, D-69117, Germany}

\altaffiltext{2}{Astronomy Department, California Institute of
  Technology, MC 105-24, 1200 East California Boulevard, Pasadena, CA
  91125; dr@caltech.edu}

\altaffiltext{3}{National Radio Astronomy Observatory, PO Box O, Socorro, NM 87801}

\altaffiltext{4}{Institute of Astronomy, School of Physics, A28,
  University of Sydney, NSW 2006, Australia}

\altaffiltext{5}{Hubble Fellow}


\begin{abstract}
  We have mapped the molecular gas content in the host galaxy of the
  strongly lensed high redshift quasar \apm\ ($z=3.911$) with the Very
  Large Array at 0.3$''$ resolution.  The \aco\ emission is clearly
  resolved in our maps. The \aco\ line luminosity derived from these
  maps is in good agreement with a previous single-dish measurement.
  In contrast to previous interferometer-based studies, we find that
  the full molecular gas reservoir is situated in two compact peaks
  separated by $\lesssim$0.4$''$. Our observations reveal, for the
  first time, that the emission from cold molecular gas is virtually
  cospatial with the optical/near-infrared continuum emission of the
  central AGN in this source.  This striking similarity in morphology
  indicates that the molecular gas is situated in a compact region
  close to the AGN. Based on the high resolution CO maps, we present a
  revised model for the gravitational lensing in this system, which
  indicates that the molecular gas emission is magnified by only a
  factor of 4 (in contrast to previously suggested factors of 100).
  This model suggests that the CO is situated in a circumnuclear disk
  of $\sim$550\,pc radius that is possibly seen at an inclination of
  $\lesssim$25$^\circ$, i.e., relatively close to face-on.  From the
  CO luminosity, we derive a molecular gas mass of $M_{\rm
  gas}$=1.3$\times$10$^{11}$\,M$_{\odot}$ for this galaxy.  From the
  CO structure and linewidth, we derive a dynamical mass of $M_{\rm
  dyn}$\,sin$^2 i$=4.0$\times$10$^{10}$\,M$_\odot$. Based on a revised
  mass estimate for the central black hole of $M_{\rm
  BH}$=2.3$\times$10$^{10}$\,M$_\odot$ and the results of our
  molecular line study, we find that the mass of the stellar bulge of
  \apm\ falls short of the local $M_{\rm BH}$--$\sigma_{\rm bulge}$
  relationship of nearby galaxies by more than an order of magnitude,
  lending support to recent suggestions that this relation may evolve
  with cosmic time and/or change toward the high mass end.
\end{abstract}

\keywords{galaxies: active --- galaxies: starburst --- galaxies: formation --- galaxies: high-redshift --- cosmology: observations 
--- radio lines: galaxies}

\section{Introduction}

Several different populations of distant galaxies have been detected
to date, out to a (spectroscopically confirmed) redshift of $z$=7.0
(see Ellis \citeyear{ell07} for a review).  It is of fundamental
importance to study such young galaxies in great detail in order to
constrain their nuclear, stellar, and gaseous constituents and their
physical properties and chemical abundances.  Understanding these
different characteristics of high redshift galaxy populations and
their progression through cosmic times is vital to develop a unified
picture of galaxy formation and evolution. One key aspect in these
high-$z$ galaxy studies are sensitive, high-resolution radio
observations of the molecular gas phase, i.e., the raw material that
fuels star formation.

Since its initial discovery at $z$$>$2 more than a decade ago (Brown
\& Vanden Bout \citeyear{bv91}; Solomon et al.~\citeyear{sol92}),
observations of spectral line emission from interstellar molecular gas
in distant galaxies have revolutionized our understanding of some of
the most luminous objects that populate the early universe. This is
due to the fact that the physical state of the molecular interstellar
medium (ISM) plays a critical role in the evolution of a galaxy. The
total amount of molecular gas in a galaxy determines for how long
starburst activity can be maintained, while its temperature and
density are a direct measure for the conditions under which star
formation can occur (see Solomon \& Vanden Bout \citeyear{sv05} for a
review).  Also, there is growing evidence that the observed
relationships between black hole mass and galaxy bulge velocity
dispersion ($M_{\rm BH}$--$\sigma_{\rm bulge}$; Ferrarese \& Merritt
\citeyear{fm2000}; Gebhardt et al.~\citeyear{g2000}), black hole mass
and host bulge concentration (i.e., Sersic index; Graham et
al.~\citeyear{gra01}; Graham \& Driver \citeyear{gd07}), and,
ultimately, black hole mass and bulge mass ($M_{\rm BH}$--$M_{\rm
  bulge}$; Magorrian et al.~\citeyear{mag98}; H\"aring \& Rix
\citeyear{hr04}), may be a consequence of an active galactic nucleus
(AGN) feedback mechanism acting on its surrounding material, at least
in the most luminous systems (Silk \& Rees \citeyear{sr98}; Di Matteo
et al.~\citeyear{mat05}). Such an AGN-driven wind will also interact
with the molecular gas, the material which will eventually form the
stellar bulges and disks in galaxies at high redshift, but also feed
the active nucleus itself. Observations of the dynamical structure and
distribution of molecular gas in distant galaxies may even reveal the
initial cause of event (e.g., mergers) for both star formation and AGN
activity, an important test for recent cosmological models (e.g.,
Springel et al.~\citeyear{spr05}; Narayanan et al.~\citeyear{nar06}).

The flux magnification provided by gravitational lensing has been
facilitated in studies of high redshift molecular gas emission since
early on, leading to the detection of a number of quasar host galaxies
(e.g., Barvainis et al.\ \citeyear{bar94}; Downes et al.\
\citeyear{dow99}) and submillimeter galaxies (SMGs; e.g., Frayer et
al.\ \citeyear{fra98}; Greve et al.\ \citeyear{gre05}), but also
substantially fainter populations such as Lyman-break galaxies (LBGs;
Baker et al.\ \citeyear{bak04}; Coppin et al.\
\citeyear{cop07}). Also, some sources have large enough lens image
separations to show structure even at only moderate angular resolution
(Sheth et al.\ \citeyear{she04}; Kneib et al.\ \citeyear{kne05}).

High angular resolution observations of molecular gas in high-$z$
galaxies are, to date, exclusively obtained in the rotational
transitions of carbon monoxide (CO). In the past few years, a number
of SMGs (Genzel et al.~\citeyear{gen03}; Downes \& Solomon
\citeyear{ds03}; Tacconi et al.~\citeyear{tac06}) and quasars (Alloin
et al.~\citeyear{all97}) at $z$$>$2 have been studied at a linear
resolution of up to 5\,kpc at the target redshifts (1\,kpc=0.12$''$ at
$z$=2). In some cases, gravitational lensing aids in zooming in
further on these systems. However, the only telescope that currently
allows to attain resolutions of sub-kpc to 1\,kpc scale at even higher
redshifts ($z$$\gtrsim$4, where 1\,kpc$\gtrsim$0.14$''$) is the NRAO's
Very Large Array (VLA)\footnote{The Very Large Array is a facility of
the National Radio Astronomy Observatory, operated by Associated
Universities, Inc., under a cooperative agreement with the National
Science Foundation.}. First sub-arcsecond resolution CO imaging with
the VLA has been obtained toward the distant quasars BR\,1202-0725
($z$=4.69; Carilli et al.~\citeyear{car02}), APM\,08279+5255
($z$=3.91; Lewis et al.~\citeyear{lew02}), PSS\,J2322+1944 ($z$=4.12;
Carilli et al.~\citeyear{car03}; Riechers et al.\ \citeyear{rie08a}),
SDSS\,J1148+5251 ($z$=6.42; Walter et al.~\citeyear{wal04}), and
BRI\,1335-0417 ($z$=4.41; Riechers et al.\ \citeyear{rie08b}).

We here report on new, more sensitive high-resolution observations of
\aco\ emission towards the strongly lensed quasar APM\,08279+5255
($z=3.911$). APM\,08279+5255 was already\footnote{Due to the source's
  enormous optical brightness, much older images may exist.}  imaged
during the Palomar Sky Survey (PSS) in 1953, a decade before the
discovery of the first quasar was reported in the literature (Schmidt
\citeyear{sch63}).  About 45\,years later, it was `officially'
discovered in an automatic plate measuring facility (APM) survey for
distant cool carbon stars, and identified as a 15$^{\rm th}$
magnitude, radio-quiet broad absorption line (BAL) quasar with an IRAS
Faint Source Catalog (FSC) counterpart at a redshift of $z_{\rm
  opt}$=3.87 (Irwin et al.~\citeyear{irw98}).  APM\,08279+5255 was
found to have an unprecedented apparent bolometric luminosity of
$L_{\rm bol}$=7$\times$10$^{15}$\,L$_\odot$, about 20\% of which is
emitted in the (far-)infrared (Lewis et al.~\citeyear{lew98}). This
extreme value was found early on to be due to strong gravitational
lensing (Irwin et al.~\citeyear{irw98}), producing three close
(maximum separation $<$0.4$''$), almost collinear images in the
optical/near-infrared (Ibata et al.~\citeyear{iba99}; Egami et
al.~\citeyear{e2000}). Due to the lack of a detection of the lensing
galaxy, and an indication for significant microlensing effects, the
true nature of the gravitational lens configuration (and thus the
magnification factor $\mu_{L}$) in this system remains subject to
debate (e.g., Egami et al.~\citeyear{e2000}; Lewis et
al.~\citeyear{lew02}) despite the fact that APM\,08279+5255 is one of
the best-studied sources in the distant universe.  In addition to its
enormous continuum brightness at basically every astronomically
relevant wavelength, APM\,08279+5255 is also one of the brightest CO
sources at high redshift (Downes et al.~\citeyear{dow99}). It also
exhibits unusually bright HCN (Wagg \etal\ \citeyear{wag05}) and
HCO$^+$ (Garcia-Burillo \etal\ \citeyear{gar06}) emission, and was
recently also detected in \ci\ (Wagg \etal\ \citeyear{wag06})
emission.  The properties of its molecular gas content have recently
been modeled based on observations of the \aco\ up to \kco\ rotational
ladder (`CO spectral line energy distribution', or SLED) with the NRAO
Green Bank Telescope (GBT)\footnote{The Green Bank Telescope is a
  facility of the National Radio Astronomy Observatory, operated by
  Associated Universities, Inc., under a cooperative agreement with
  the National Science Foundation.} and the IRAM 30\,m telescope
(Riechers et al.~\citeyear{rie06}; Wei\ss\ et al.~\citeyear{wei07}).
We use a standard concordance cosmology throughout, with $H_0 =
71\,$\kms\,Mpc$^{-1}$, $\Omega_{\rm M} =0.27$, and $\Omega_{\Lambda} =
0.73$ (Spergel \etal\ \citeyear{spe03}, \citeyear{spe07}).

\section{Observations}

We here report new, sensitive high-resolution observations of the
\aco\ emission line in \apm. To better constrain the molecular gas
properties in this source, we also re-derive the strength of the
\bco\ emission line, based on more sensitive (unpublished archival) 
observations than reported previously (Papadopoulos et al.\
\citeyear{pap01}). In this context, we also present a search for the
\acn\ emission line in this galaxy, a tracer of the dense fraction of
the molecular gas phase (see Riechers \etal\ \citeyear{rie06} for more
details on this data).  To further investigate the AGN and starburst
environments in this source, we augment our analysis of the molecular
gas properties with a study of the radio continuum properties, based
in part on archival observations. A fraction of the archival continuum
data was used in previous studies of this source (Ibata et al.\
\citeyear{iba99}; Ivison \citeyear{ivi06}; see Tab.~\ref{t1} for
details).

\begin{deluxetable}{lccccc}
\tabletypesize{\scriptsize}
\tablecaption{New (2005) and archival (1998--2001)\\ VLA observations summary. \label{t1}}
\tablehead{
Date & Config.\ & Duration & Band & Type & Ref.\ }
\startdata
2005 Apr 07 & B   & 6.75\,h & K & \aco      & 1 \\
2005 Apr 08 & B   & 6.0\,h  & K & \aco      & 1 \\
2005 Apr 09 & B   & 5.25\,h & K & \aco      & 1 \\
2005 Apr 10 & B   & 6.0\,h  & K & \aco      & 1 \\
2005 Apr 18 & B   & 7.5\,h  & K & \aco      & 1 \\
2005 May 01 & B   & 4.0\,h  & C & Continuum & 1 \\
2005 Sep 12 & C   & 8.25\,h & K & \aco      & 1 \\
2005 Sep 20 & C   & 7.5\,h  & K & \aco      & 1 \\
2005 Dec 02 & D   & 4.0\,h  & K & \aco      & 1 \\
\tableline
1998 Jun 18 & BnA & 3.0\,h  & X & Continuum & 2 \\
1998 Jul 28 & B   & 2.5\,h  & X & Continuum & 1 \\
            &     & 2.0\,h  & U & Continuum & 1 \\
1998 Aug 11 & B   & 3.0\,h  & X & Continuum & 1 \\
            &     & 3.5\,h  & U & Continuum & 1 \\
2000 Apr 24 & C   & 13.0\,h & Q & \bco      & 3 \\
2000 Sep 16 & D   & 4.0\,h  & Q & \bco      & 1 \\
2001 Jan 21 & A   & 2.25\,h  & L & Continuum & 4 \\
2001 Oct 20 & D   & 9.0\,h  & Q & \bco      & 1 \\
\vspace{-3mm}
\enddata
\tablerefs{[1] this work; [2] Ibata et al.\ \citeyear{iba99}; [3] Papadopoulos et al.\ \citeyear{pap01}; [4] Ivison \citeyear{ivi06}.}
\tablecomments{Maximum baselines of VLA configurations are: A - 36.4\,km, B - 11.4\,km, C - 3.4\,km, D - 1.03\,km. BnA is a hybrid configuration where the antennas on the east and west arms of the array are in the shorter of the two denoted configurations, and the north arm is in the more extended configuration.\\ }
\end{deluxetable}

\subsection{VLA Data} \label{newobs}

\subsubsection{\aco\ and 23.5\,GHz continuum [new]}

We observed the \aco\ transition line ($\nu_{\rm rest} =
115.2712018\,$GHz) towards APM\,08279+5255 using the VLA in B, C, and
D configuration between 2005 April 07 and 2005 December 02. At the
target redshift of 3.911, the line is redshifted to 23.472\,GHz
(12.77\,mm, K band).  The total on-sky integration time amounts to
51.25\,hr (see Tab.~\ref{t1} for details).  Observations were
performed in fast-switching mode (see, e.g., Carilli \& Holdaway
\citeyear{ch99}) using the nearby source 08248+55527 for secondary
amplitude and phase calibration.  Observations were carried out under
excellent weather conditions with 25 antennas.  The phase stability in
all runs was excellent (typically $<$15$^\circ$ for the longest
baselines).  The phase coherence was determined by imaging the
calibrator source 08087+49506 with the same calibration cycle as that
used for the target source, and found to be high during all runs.  For
primary flux calibration, 3C\,286 was observed during each run.

Two 25\,MHz intermediate frequencies (IFs) with seven 3.125\,MHz
channels each were observed in `spectral line' mode centered at the
\aco\ line frequency, leading to an effective bandwidth of 43.75\,MHz
(corresponding to 558\kms\ at 23.5\,GHz; `ON' channels).  Two 50\,MHz
(corresponding to 638\kms\ at 23.5\,GHz) IFs were observed in
continuum mode at 23.3649\,GHz and 23.5649\,GHz ($\pm$100\,MHz offset
from the line frequency) to measure the source's continuum at 12.8\,mm
(`OFF' channels).  The continuum was observed for one third of the
total time to attain the same rms as in the combined line channels.

\begin{figure*}
\epsscale{1.15}
\vspace{-3mm}

\plotone{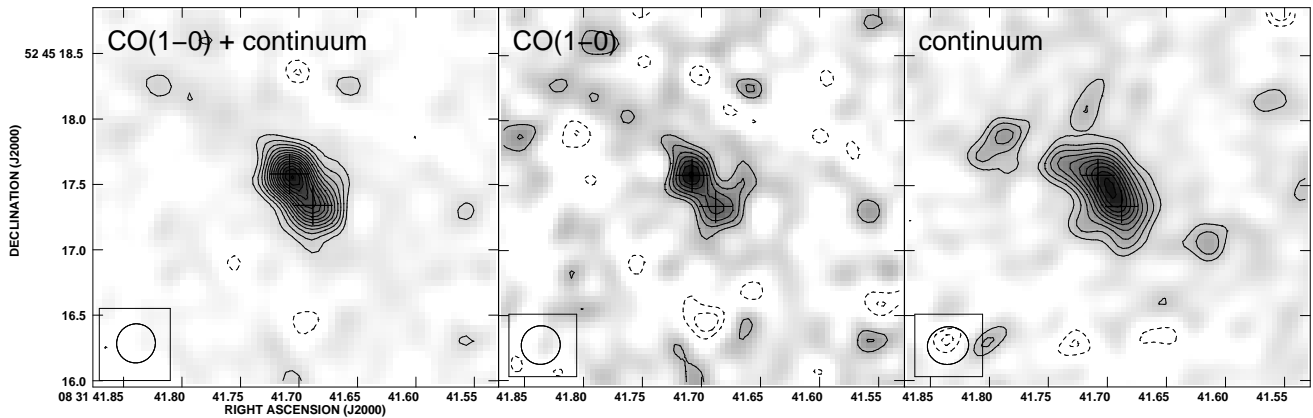}
\vspace{-8mm}

\caption{High-resolution VLA maps of \aco\ + continuum ({\em left}), \aco\ only ({\em
  middle}), and 23.5\,GHz continuum ({\em right}) emission towards
  APM\,08279+5255 at a resolution of
  0.30/0.30/0.32\,$''$$\times$0.29\,$''$ (robust 1 weighting; left to
  right; beam size indicated in the bottom left corner of each panel).
  Contours are shown in steps of the rms noise, starting at
  $\pm$2$\sigma$ (1$\sigma = 16/16/13\,\mu$Jy beam$^{-1}$).  {\em
  Left:} Image of the \aco\ and continuum emission.  The source is
  clearly resolved, and detected at a peak strength of 15$\sigma$.
  This map is integrated over the central 558\kms\ (43.75\,MHz) of the
  \aco\ emission.  {\em Middle:} Image of the \aco\ emission.
  Continuum emission was subtracted.  The source is clearly resolved
  into two CO peaks (8$\sigma$ and 5$\sigma$ signal-to-noise), as
  indicated by the plus signs.  This continuum-subtracted map is
  integrated over the same range as in {\em left} the panel.  {\em
  Right:} Image of the 23.5\,GHz continuum emission (peak:
  11$\sigma$).  The emission appears resolved and peaks centrally
  between the \aco\ peaks (as indicated by the plus signs).  This
  continuum map is integrated over a bandwidth of 100\,MHz.
\label{f1}}
\end{figure*}

For data reduction and analysis, the
${\mathcal{AIPS}}$\footnote{www.aoc.nrao.edu/aips/} package was used.
The two continuum channels were concatenated in the uv/visibility
plane.  The \aco\ line image was generated by subtracting a CLEAN
component model of the continuum emission from the visibility data.
All data were mapped using the CLEAN algorithm. The
velocity-integrated maps shown in Fig.\ \ref{f1} are imaged using
robust 1 weighting and combining B and C array data, leading to a
synthesized clean beam size of 0.30\,$''$$\times$0.29\,$''$ in the
\aco\ line map and 0.32\,$''$$\times$0.29\,$''$ in the continuum map
(note that D array data were omitted in these maps to improve the size
of the synthesized beam).  The final rms over the full bandwidth of
43.75\,MHz (558\,\kms) in the \aco\ map is 16\,$\mu$Jy beam$^{-1}$,
and the rms over the 100\,MHz continuum bandpass is 13\,$\mu$Jy
beam$^{-1}$.  In Fig.\ \ref{f3}, four velocity channel maps
(9.375\,MHz, or 120\,\kms\ each) of the central 37.5\,MHz (480\,\kms)
of the \aco\ line are shown. To maximize the sensitivity in individual
channels the data were imaged using natural weighting and combining B
and C array data, achieving a resolution of
0.34\,$''$$\times$0.33\,$''$.  Hanning smoothing was applied to the
image cube, resulting in an rms of 27\,$\mu$Jy\,beam$^{-1}$ per
channel.  The velocity-integrated \aco\ and 23.5\,GHz continuum maps
of the combined B, C {\em and D} array data shown in Fig.\ \ref{f5}
and Fig.\ \ref{f15}, top right, are imaged using natural weighting and
a Gaussian taper, leading to synthesized beams of
1.19\,$''$$\times$1.14\,$''$ and 1.19\,$''$$\times$1.10\,$''$, and an
rms of 19\,$\mu$Jy beam$^{-1}$ for both maps.

\subsubsection{\bco\ [archival]}

Spectral line observations of the \bco\ transition line ($\nu_{\rm
  rest} = 230.53799\,$GHz) towards APM\,08279+5255 were carried out
between 2000 Apr 24 and 2001 Oct 20 using the VLA in C and D
configuration. The C array data was published by Papadopoulos et al.\
(\citeyear{pap01}). We here re-analyze this data, and add in shorter
(unpublished) D array spacings to improve the sensitivity of the data.
At the target redshift of 3.911, the \bco\ line is redshifted to
46.943\,GHz (6.39\,mm, Q band).  The total on-sky integration time
amounts to 26\,hr (see Tab.~\ref{t1} for details).  Observations were
performed in fast-switching mode using the nearby source 08248+55527
for secondary amplitude and phase calibration.  Observations were
carried out under excellent weather conditions with up to 27 antennas.
For primary flux calibration, 3C\,48 and 3C\,286 were observed.
Observations were set up in quasi-continuum mode; two 50\,MHz IFs
were observed simultaneously at 46.9149\,GHz and 46.9649\,GHz to cover
the central $\sim$640\kms\ of the spectral line.

Data were mapped using the CLEAN algorithm at natural weighting; this
results in a synthesized beam of 1.11\,$''$$\times$0.93\,$''$ and an
rms of 65\,$\mu$Jy beam$^{-1}$ (no map shown here).

\begin{deluxetable}{lcc}
\tabletypesize{\scriptsize}
\tablecaption{CO/CN Line Peak and Continuum Fluxes. \label{t2}}
\tablehead{
Transition/Frequency & $S_{\nu}$ & Configuration\\
& [$\mu$Jy] & }
\startdata
\aco & 285$\pm$27\tablenotemark{a} & BCD \\
\bco & 1637$\pm$65\tablenotemark{b} & CD \\
\acn & $<$195 & [GBT] \\
\tableline
1.4\,GHz  & 1160$\pm$33 & A\\
4.5\,GHz & 551$\pm$40 & B \\
8.4\,GHz & 446$\pm$20 & BnA/B\\
14.9\,GHz & 303$\pm$93\tablenotemark{c} & B\\
23.5\,GHz & 376$\pm$19 & BCD\\
46.9\,GHz (estimated)\tablenotemark{d} & 405$^{+310}_{-330}$ & ---\\
\vspace{-2.5mm}
\enddata
\tablenotetext{a}{Peak flux from Gaussian fitting to the unsmoothed 14 channel spectrum.} 
\tablenotetext{b}{Not corrected for the estimated 46.9\,GHz continuum contribution.}
\tablenotetext{c}{Tentative detection.}
\tablenotetext{d}{Estimated based on the FIR-to-radio SED. The lower limit is given by the model of the FIR-only part of the continuum, and the upper limit is estimated by linear interpolation (in the log--log plane) of the 23\,GHz and 94\,GHz continuum measurements.}
\tablecomments{Fluxes are derived from convolved maps for all datasets with linear resolutions higher than 1.2$''$, and agree with those derived from Gaussian fitting to the unconvolved data within 10\%.}
\end{deluxetable}

\subsubsection{1.4\,GHz continuum [archival]}

Continuum observations at 1.400\,GHz (L band) towards APM\,08279+5255
were carried out on 2001 Jan 21 using the VLA in A configuration.
This data was recently published by Ivison (\citeyear{ivi06}; no map
shown).  The total on-sky integration time amounts to 2.25\,hr (see
Tab.~\ref{t1} for details).  Observations were performed in spectral
line mode using the nearby source 0824+558 for secondary amplitude and
phase calibration.  Observations were set up using two IFs and seven
3.125\,MHz channels per IF, leading to an effective bandwidth of
43.75\,MHz (corresponding to $\sim$9400\,\kms\ at 1.4\,GHz).
Observations were carried out under good weather conditions with 27
antennas. For primary flux calibration, 3C\,48 was observed.

Data were mapped using the CLEAN algorithm at robust 0 weighting; this
results in a synthesized beam of 1.42\,$''$$\times$1.15\,$''$ and an
rms of 33\,$\mu$Jy beam$^{-1}$ (see Fig.\ \ref{f15}, top left).

\subsubsection{4.5\,GHz continuum [new]}

Continuum observations at 4.5276\,GHz (C band) towards APM\,08279+5255
were carried out on 2005 May 01 using the VLA in B configuration. The
total on-sky integration time amounts to 4\,hr (see Tab.~\ref{t1} for
details).  Observations were performed in spectral line mode using the
nearby source 08248+55527 for secondary amplitude and phase
calibration.  Observations at 4.5276\,GHz were set up using thirty-one
390.625\,kHz channels, leading to an effective bandwidth of
12.109375\,MHz (corresponding to 802\kms\ at 4.5276\,GHz).
Observations were carried out under excellent weather conditions with
25 antennas. The phase stability in this run was excellent.  For
primary flux calibration, 3C\,286 was observed during each run.

Data were mapped using the CLEAN algorithm at natural weighting; this
results in a synthesized beam of 1.55\,$''$$\times$1.42\,$''$ and an
rms of 40\,$\mu$Jy beam$^{-1}$ (see Fig.\ \ref{f15}, bottom left).

\subsubsection{8.4\,GHz continuum [archival]}

Continuum observations at 8.4\,GHz (X band) towards APM\,08279+5255
were carried out between 1998 Jun 18 and Aug 11 using the VLA in BnA
and B configuration. The BnA array data was published by Ibata et al.\
(\citeyear{iba99}). We here re-analyze this data, and add in shorter B
array spacings to improve the sensitivity and image quality of the
data. The total on-sky integration time amounts to 8.5\,hr (see
Tab.~\ref{t1} for details).  Observations were performed in
quasi-continuum mode (two 50\,MHz IFs were observed simultaneously at
8.4351\,GHz and 8.4851\,GHz) using the nearby sources 0749+540 and
0820+560 for secondary amplitude and phase calibration. Observations
were carried out under good weather conditions with 27 antennas. For
primary flux calibration, 3C\,48 and 3C\,147 were observed.

Data were mapped using the CLEAN algorithm. Imaging the data with
natural weighting results in a synthesized beam of
0.73\,$''$$\times$0.42\,$''$ and an rms of 10\,$\mu$Jy beam$^{-1}$
(see Fig.\ \ref{f15}, bottom right). Imaging the data with uniform
weighting results in a synthesized beam of
0.63\,$''$$\times$0.22\,$''$ and an rms of 20\,$\mu$Jy beam$^{-1}$
(see Fig.~\ref{f9}).

\subsubsection{14.9\,GHz continuum [archival]}

Continuum observations at 14.9\,GHz (U band) towards APM\,08279+5255
were carried out on 1998 Jul 28 and Aug 11 using the VLA in B
configuration. The total on-sky integration time amounts to 5.5\,hr
(see Tab.~\ref{t1} for details).  Observations were performed in
quasi-continuum mode (two 50\,MHz IFs were observed simultaneously at
14.9149\,GHz and 14.9649\,GHz) using the nearby source 0749+540 for
secondary amplitude and phase calibration.  Observations were carried
out under good weather conditions with 27 antennas. For primary flux
calibration, 3C\,48 was observed.

Data were mapped using the CLEAN algorithm at natural weighting; this
results in a synthesized beam of 0.51\,$''$$\times$0.44\,$''$ and an
rms of 44\,$\mu$Jy beam$^{-1}$ (no map shown here).

\begin{figure*}
\vspace{-11mm}

\epsscale{1.2}
\plotone{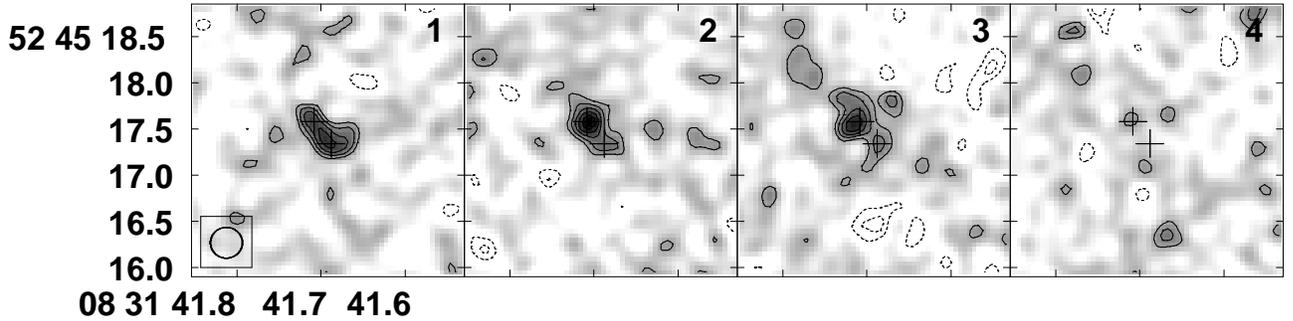}
\vspace{-28mm}

\caption{
Channel maps of the \aco\ emission (same region is shown as in
Fig.~\ref{f1}). The beam size (0.34\,$''$$\times$0.33\,$''$; natural
weighting) is shown in the bottom left corner. The plus signs indicate
the same positions as in Fig.~\ref{f1}.  Contours are shown in steps
of the rms noise, starting at $\pm$2$\sigma$ (1$\sigma = 27\,\mu$Jy
beam$^{-1}$). One channel width is 9.375\,MHz, or 120\,km\,s$^{-1}$
(frequencies increase with channel number and are shown at 23451.539,
23460.914, 23470.289, and 23479.664\,MHz).  \label{f3}}
\end{figure*}

\subsection{GBT Data} \label{gbtobs}

\subsubsection{CN($N$=1$\to$0)}

As part of our wide-band high-$z$ molecular line study (Riechers et
al.\ \citeyear{rie06}), we observed the $F$=3/2$\to$1/2,
$F$=5/2$\to$3/2, $F$=1/2$\to$1/2, and $F$=3/2$\to$3/2 hyper-fine
structure transitions of the CN($N$=1$\to$0, $v$=0, $J$=3/2$\to$1/2)
line ($\nu_{\rm rest} = 113.4881420-113.5089340\,$GHz) towards
APM\,08279+5255 with the GBT between 2004 December 15 and 19 for a
total of 22.5\,hr (15\,hr on source).  At the target redshift of
3.911, these lines are redshifted to 23.109-23.113\,GHz (12.97\,mm, K
band), and can be observed simultaneously (in-band) with \aco\ at the
GBT (see Riechers et al.\ \citeyear{rie06} for a more detailed
description and results on the CO observations).  Observations were
performed in ON--OFF position switching mode using the nearby source
J0753+538 for secondary amplitude and pointing calibration, correcting
for the atmospheric opacity. The pointing accuracy was typically
3$''$.  Observations were carried out under excellent weather
conditions.  For primary flux calibration, 3C\,147 and 3C\,286 were
observed during each run.  The dual-beam, dual-polarization 18-26\,GHz
receiver was used for all observations.  The beam size at our
observing frequency is 32$''$ ($\sim$230\,kpc at the source's
redshift), i.e., much larger than our target.

Two 800\,MHz IFs with 2048 390.625\,kHz (5\,\kms) channels and two
orthogonal polarizations each were observed simultaneously. The first
IF was centered on the redshifted \aco\ frequency at 23.4663\,GHz, and
the second IF was tuned 700\,MHz bluewards. Therefore, the observed
\acn\ fine structure transition lines are located at the lower end of
the first IF.

For data reduction and analysis, the
AIPS$^{++}$\footnote{http://www.aips2.nrao.edu.} and
GBT IDL\footnote{http://gbtidl.sourceforge.net.} packages were used.
Final combination, binning, and baseline subtraction were done with
GILDAS/CLASS\footnote{http://www.iram.fr/IRAMFR/GILDAS/.}.
An rms of 65\,$\mu$Jy was attained in these observations.

\section{Results}

\subsection{CO(1--0) Line and 23.5\,GHz Continuum Emission} \label{aco}

We have resolved the emission from the \aco\ transition line towards
APM\,08279+5255 ($z=3.91$), as well as the continuum emission at the
same frequency (Fig.~\ref{f1}). The left panel of Fig.~\ref{f1} shows
the \aco\ and continuum emission, the middle panel the
continuum-subtracted \aco\ emission, and the right panel the 23.5\,GHz
continuum emission.  The continuum-subtracted \aco\ line emission
(Fig.\ \ref{f1}, middle) is clearly resolved into two peaks
(north-east and south-west).  From 2-dimensional Gaussian fitting we
find a line peak flux density of 131 $\pm$ 16\,$\mu$Jy beam$^{-1}$ for
the north-eastern CO peak at $\alpha$=$08^{\rm h}31^{\rm m}41^{\rm
s}.708$ $\pm$ $0^{\rm s}.001$, $\delta$=$+52^\circ45'17''.58$ $\pm$
$0''.02$ and a line peak flux density of 87 $\pm$ 16\,$\mu$Jy
beam$^{-1}$ for the south-western CO peak at $\alpha$=$08^{\rm
h}31^{\rm m}41^{\rm s}.688$ $\pm$ $0^{\rm s}.004$,
$\delta$=$+52^\circ45'17''.34$ $\pm$ $0''.05$ (astrometric errors are
derived from the fit; Tables \ref{t2} and \ref{t2a}). This corresponds
to a peak flux ratio of $b$(SW,NE)=0.66 $\pm$ 0.15. The separation
between both peaks is $d$(SW,NE)=0.31$''$ $\pm$ 0.05$''$.

The continuum emission (Fig.\ \ref{f1}, right) is also clearly
resolved and has about the same extent as the \aco\ line emission, but
is not resolved into two individual peaks. The emission rather peaks
centrally between the two CO peaks indicated by the plus signs.  For
the continuum emission at this resolution, we derive a peak flux
density of 147 $\pm$ 12\,$\mu$Jy beam$^{-1}$ by fitting a single
two-dimensional Gaussian.

In Fig.\ \ref{f3}, four (continuum-subtracted) channel maps
(9.375\,MHz, or 120\,\kms\ each) of the central 37.5\,MHz (480\,\kms)
of the \aco\ line are shown.  The line is resolved into the two \aco\
peaks defined above in the central two channels, and the decline of
the line intensity towards the line wings is clearly visible in the
outer channels, as expected.  Due to the slight asymmetry of the \aco\
line with respect to our central tuning frequency, only little
emission is detected in the blueward outer channel, the lack of
extended emission in this channel is therefore ascribed to low
signal-to-noise rather than different structure relative to the other
channels.  However, there is tentative evidence for a spatially
resolved velocity gradient between the velocity channels along an axis
close to the major lensing axis. From the central line channels, we
derive a line peak flux density of 185 $\pm$ 27\,$\mu$Jy beam$^{-1}$
for the north-eastern CO peak and a line peak flux density of 101
$\pm$ 27\,$\mu$Jy beam$^{-1}$ for the south-western CO peak.

\begin{figure}
\vspace{-4mm}

\epsscale{1.15}
\plotone{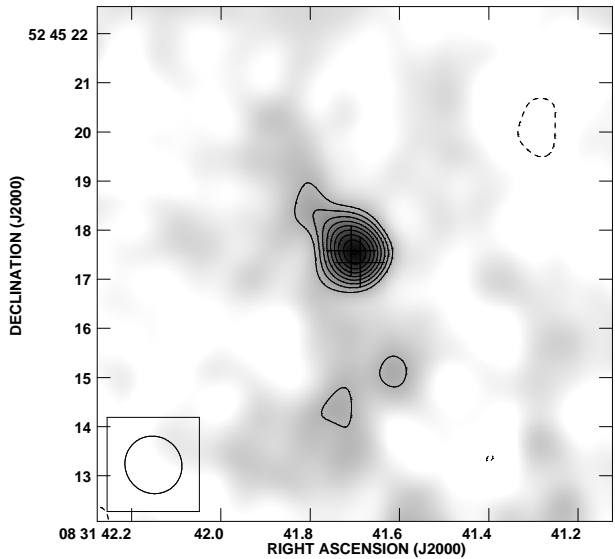}
\vspace{-9.5mm}

\caption{
Map of the \aco\ emission as shown in Fig.~\ref{f1} ({\em middle}),
but including D array short spacings, and tapered to a resolution of
1.19\,$''$$\times$1.14\,$''$ (natural weighting; beam size indicated
in the bottom left corner).  Contours are shown in steps of the rms
noise, starting at $\pm$3$\sigma$ (1$\sigma = 19\,\mu$Jy beam$^{-1}$;
11$\sigma$ peak). The plus signs indicate the same positions as in
Fig.~\ref{f1}.  The emission appears practically unresolved, and there
is no indication of any extended emission.
\label{f5}}
\vspace{2mm}

\end{figure}

To derive the full \aco\ and 23.5\,GHz continuum flux of
APM\,08279+5255, we have imaged the whole B, C, and D array dataset
with natural weighting, but tapered to a linear resolution of
$\sim$1.2$''$ (Fig.\ \ref{f5} and Fig.\ \ref{f15}, top right).  At
this resolution, the source appears at best marginally resolved, and
we derive that we outresolve at most 10\% of the integrated flux.
From the integrated maps, we derive an \aco\ peak flux density of 210
$\pm$ 19\,$\mu$Jy beam$^{-1}$, and a continuum peak flux density of
372 $\pm$ 19\,$\mu$Jy beam$^{-1}$. The complete beam-corrected source
fluxes (derived from 2-dimensional Gaussian fitting to the source) in
both maps agree with the peak fluxes within the errors. It also agrees
well with the beam-corrected fluxes in the full resolution dataset
within the error bars. From the central line channels, we derive a
\aco\ line peak flux density of 285 $\pm$ 27\,$\mu$Jy beam$^{-1}$.

This peak flux agrees well with that derived from single-dish
observations. Figure~\ref{f4} shows the central part of the \aco\
spectrum we have obtained with the GBT (Riechers et al.\
\citeyear{rie06}). This spectrum has a resolution of 75\,\kms\ at an
rms of 65\,$\mu$Jy per channel. The solid black line is a Gaussian fit
to that data. Overplotted on this spectrum, we show the flux of the
four 120\,\kms\ channels of Fig.\ \ref{f3} (but for the tapered BCD
array dataset).  The error bars indicate the rms per channel
(40\,$\mu$Jy). This clearly demonstrates that, within the error bars,
our interferometric observations recover the same amount of CO
emission as the single-dish observations obtained with a $\sim$32$''$
observing beam.

\subsection{CO(2--1) Line Emission} \label{bco}

The \bco\ transition line towards APM\,08279+5255 ($z=3.91$) is also
detected (see also discovery paper by Papadopoulos et al.\
\citeyear{pap01}).  To extract the full \bco\ line flux, we convolved
the map with a circular beam of 1.2$''$ diameter.  From 2-dimensional
Gaussian fitting, we then derive a line peak flux density of 1637
$\pm$ 65\,$\mu$Jy beam$^{-1}$.

We note that during none of the (archival) \bco\ observations, the
continuum at 46.9\,GHz was measured. However, Papadopoulos et al.\
(\citeyear{pap01}) report a limit of 3$\sigma$=0.9\,mJy on the
continuum emission at 43.3\,GHz.  This means that up to about half of
the \bco\ emission reported above may be attributed to 7\,mm continuum
emission rather than \bco\ emission (see Sect.~4.3 for further
discussion).

\subsection{The Search for CN(1--0) Line Emission} \label{acn}

To measure the total CO luminosity, we have obtained wide-band \aco\
observations toward \apm\ with the GBT (see Riechers et al.\
\citeyear{rie06} for details). The bandwidth of these observations was
high enough to also search for redshifted \acn\ emission within the
same setup (see Fig.~\ref{f14}). The strongest hyper-fine structure
components of the CN($N$=1$\to$0, $v$=0, $J$=3/2$\to$1/2) multiplet,
i.e.\ the $F$=3/2$\to$1/2, $F$=5/2$\to$3/2, $F$=1/2$\to$1/2, and
$F$=3/2$\to$3/2 transitions, have rest frequencies of 113.4881420,
113.4909850, 113.4996430, and 113.5089340\,GHz. At $z=3.911$, this
corresponds to 23.109, 23.110, 23.111, and 23.113\,GHz. Accounting for
the uncertainty in redshift (Riechers et al.\ \citeyear{rie06}), the
dashed lines in Fig.~\ref{f14} indicate the region where these
transitions are expected to peak. We note that, if the CN emission
comes from a similar region than the CO emission, these components are
expected to be blended due to their large linewidth
($\gtrsim$500\,\kms, or 39\,MHz FWHM, Riechers et al.\
\citeyear{rie06}; Wei\ss\ et al.\ \citeyear{wei07}).  At an rms of
65\,$\mu$Jy, we do not detect any emission. We thus set a 3$\sigma$
limit of 195\,$\mu$Jy on the peak strength of the \acn\ emission in
APM\,08279+5255.

\begin{figure*}
\epsscale{1.15}
\vspace{-4mm}

\plotone{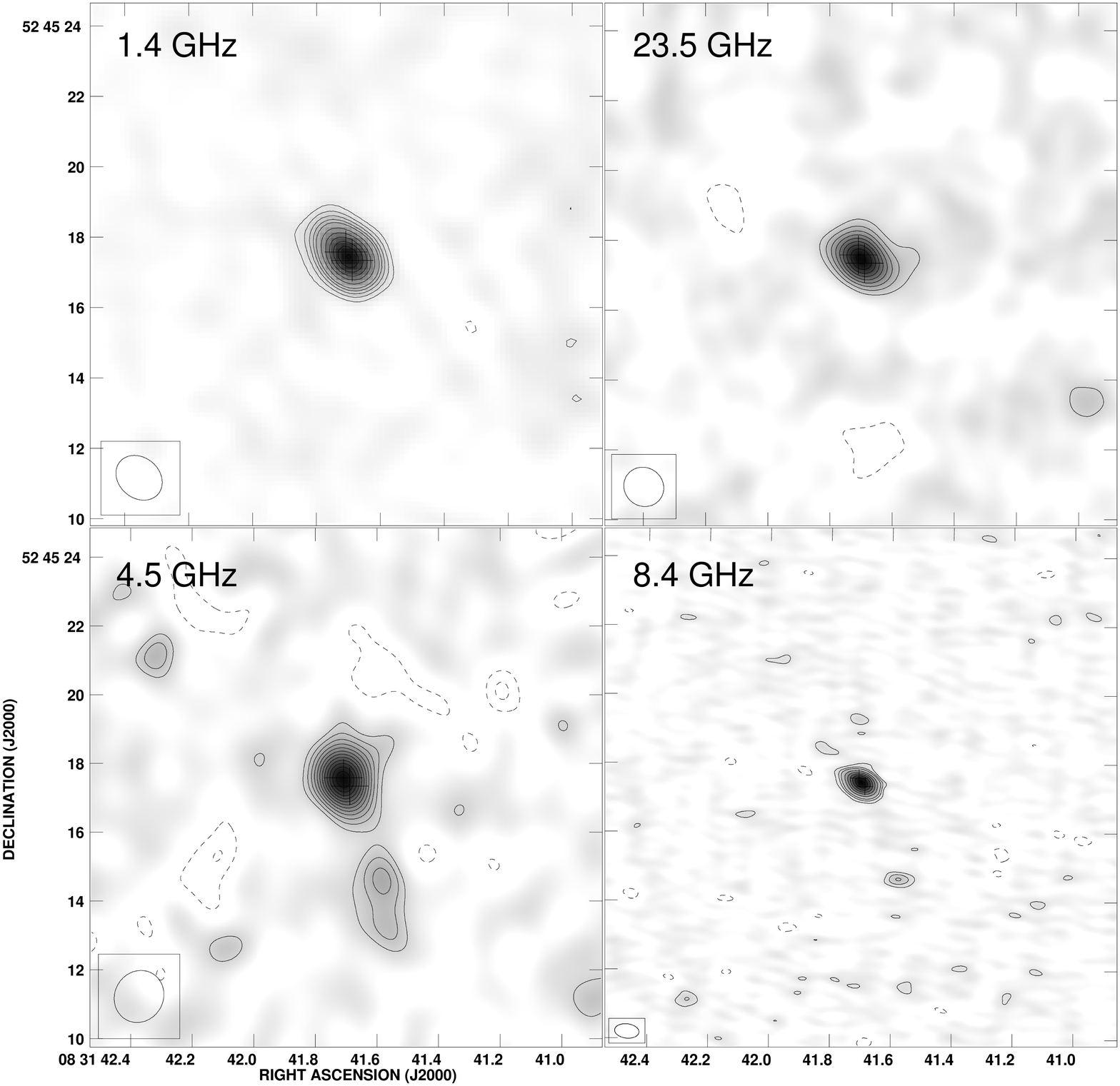}
\vspace{-11.5mm}

\caption{
VLA maps of 1.4, 4.5, 8.4, and 23.5\,GHz continuum emission towards
APM\,08279+5255 (1.4\,GHz: robust 0, rest: natural weighting). The
plus signs indicate the same positions as in Fig.~\ref{f1}.  {\em Top
left:} Image of the 1.4\,GHz continuum emission at a resolution of
1.42\,$''$$\times$1.15\,$''$ (as indicated in the bottom left corner).
Contours are shown at (-3, 3, 5, 7, 9, 11, 15, 19, 23, 27, 31,
35)$\times\sigma$ (1$\sigma = 33\,\mu$Jy beam$^{-1}$).  This continuum
map is integrated over a bandwidth of 43.75\,MHz. {\em Top right:}
Image of the 23.5\,GHz continuum emission tapered to a resolution of
1.19\,$''$$\times$1.10\,$''$. Contours are shown in steps of twice the
rms noise, starting at $\pm$3$\sigma$ (1$\sigma = 19\,\mu$Jy
beam$^{-1}$; 19$\sigma$ peak).  This continuum map is integrated over
a bandwidth of 100\,MHz.  {\em Bottom left:} Image of the 4.5\,GHz
continuum emission at a resolution of 1.55\,$''$$\times$1.42\,$''$.
Contours are shown in steps of the rms noise, starting at
$\pm$2$\sigma$ (1$\sigma = 40\,\mu$Jy beam$^{-1}$; 13$\sigma$
peak). This continuum map is integrated over a bandwidth of
12.11\,MHz.  {\em Bottom right:} Image of the 8.4\,GHz continuum
emission at a resolution of 0.73\,$''$$\times$0.42\,$''$.  Contours
are shown at (-3, 3, 5, 7, 11, 15, 19, 23, 27, 31)$\times\sigma$
(1$\sigma = 10\,\mu$Jy beam$^{-1}$). This continuum map is integrated
over a bandwidth of 100\,MHz. In the 4.5 and 8.4\,GHz maps, a radio
source is detected about 3$''$ south of APM\,08279+5255 (at
$\alpha=08^{\rm h}31^{\rm m}41^{\rm s}.576$,
$\delta=+52^\circ45'14''.66$; 4$\sigma$ and 7$\sigma$ peak strength).
This source remains undetected at 1.4 and 23.5\,GHz (also undetected
at 14.9 and 46.9\,GHz).
\label{f15}}
\end{figure*}

\begin{deluxetable*}{rccc}
\tabletypesize{\scriptsize}
\tablecaption{Spatially Resolved Components. \label{t2a}}
\tablehead{
Component   & Position             & $S_{\nu}$(CO) & $S_{\nu}$(8.4\,GHz) \\ 
            &  ($\alpha$;$\delta$) & [$\mu$Jy]     & [$\mu$Jy] }
\startdata
APM08279(A) & ($08^{\rm h}31^{\rm m}41^{\rm s}.708$ $\pm$ $0^{\rm s}.001$;$+52^\circ45'17''.58$ $\pm$ $0''.02$) & 131 $\pm$ 16 & 205 $\pm$ 20 \\
        (B) & ($08^{\rm h}31^{\rm m}41^{\rm s}.688$ $\pm$ $0^{\rm s}.004$;$+52^\circ45'17''.34$ $\pm$ $0''.05$) & 87 $\pm$ 16 & 204 $\pm$ 20 \\
\vspace{-3mm}
\enddata
\end{deluxetable*}

\subsection{Continuum Emission at 1.4\,GHz} \label{1.4ghz}

Using the VLA in A array, 1.4\,GHz continuum emission from
APM\,08279+5255 was detected by Ivison (\citeyear{ivi06}).  From this
analysis, Ivison (\citeyear{ivi06}) derived a continuum flux density
of 3.05 $\pm$ 0.07\,mJy. APM\,08279+5255 was previously detected at
1.4\,GHz in the VLA-FIRST survey (White et al.~\citeyear{whi97}), but
at significantly lower resolution and signal-to-noise.  However, the
1.4\,GHz continuum flux of 897 $\pm$ 148\,$\mu$Jy beam$^{-1}$ derived
from the VLA-FIRST data is significantly lower that that of Ivison
(\citeyear{ivi06}). To obtain a coherent dataset, we thus re-analyzed
the A-array data by Ivison (\citeyear{ivi06}).  We derive a continuum
peak flux density of 1160 $\pm$ 33\,$\mu$Jy beam$^{-1}$ (see Fig.\
\ref{f15}, top left). This value is consistent with that found from
the VLA-FIRST survey data, but significantly lower than that derived
by Ivison (\citeyear{ivi06}). Our result was independently confirmed
by Ivison (2007, priv.~comm.).  We thus use our revised 1.4\,GHz
continuum flux in the following.

\subsection{Continuum Emission at 4.5\,GHz} \label{4.5ghz}

We have detected continuum emission at 4.5\,GHz towards
APM\,08279+5255 using the VLA in B array.  APM\,08279+5255 remains
unresolved in our observations (see Fig.\ \ref{f15}, bottom left).  We
derive a continuum peak flux density of 551 $\pm$ 40\,$\mu$Jy
beam$^{-1}$.  We also detect another continuum source about 3$''$
south of APM\,08279+5255, which is detected at a peak flux density of
180 $\pm$ 40\,$\mu$Jy beam$^{-1}$ (this source was first detected at
8.4\,GHz by Ibata et al.\ \citeyear{iba99}, see peak position derived
below). At the given signal-to-noise, it is unclear whether this
source is extended or not.  APM\,08279+5255 has been detected at
4.5\,GHz previously (Ivison \citeyear{ivi06}), but at a factor of 4
worse linear resolution (VLA C array, $\sim 5.5''$).  Thus,
APM\,08279+5255 and the source detected 3$''$ south are blended at
this resolution in Ivison's data, and the continuum flux of
APM\,08279+5255 at 4.5\,GHz is overestimated. The 4.5\,GHz continuum
flux derived from Ivison's data (his Fig.~2, bottom panel) is in good
agreement with the sum of the 4.5\,GHz continuum fluxes of both
sources in our analysis.  We thus use our revised 4.5\,GHz continuum
flux of \apm\ in the following.

\subsection{Continuum Emission at 8.4\,GHz} \label{8.4ghz}

\begin{figure}
\vspace{-4mm}

\epsscale{1.15}
\plotone{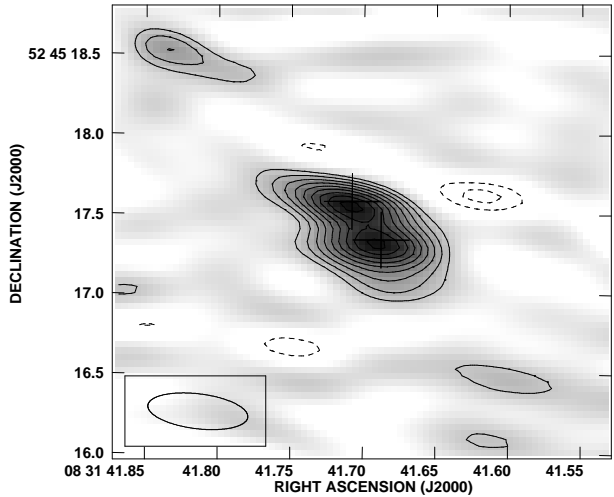}
\vspace{-9mm}

\caption{
Uniformly weighted 8.4\,GHz continuum map.  The beam size
(0.63\,$''$$\times$0.22\,$''$) is shown in the bottom left corner.
Contours are shown in steps of the rms noise, starting at
$\pm$2$\sigma$ (1$\sigma = 20\,\mu$Jy beam$^{-1}$).  The source is
clearly resolved into two peaks at this resolution (10$\sigma$ each),
which coincide with the \aco\ peaks (as indicated by the plus signs,
see Fig.~\ref{f1}).
\label{f9}}
\end{figure}

Continuum emission from APM\,08279+5255 was first detected at 8.4\,GHz
by Ibata et al.\ (\citeyear{iba99}) using the VLA in BnA array. We
here combine this data with shorter B array spacings to improve the
sensitivity of the data (see Fig.\ \ref{f15}, bottom right).  For
APM\,08279+5255, we derive a 8.4\,GHz peak flux density of 446 $\pm$
20\,$\mu$Jy beam$^{-1}$ (map convolved with a circular beam of 1.2$''$
diameter for flux extraction), consistent with the value given by
Ibata et al.\ (\citeyear{iba99}) of 0.45 $\pm$ 0.03\,mJy. We also
detect the `southern' source described in the previous subsection. It
is detected at higher signal-to-noise and resolution in the
unconvolved map than at 4.5\,GHz. We thus use the 8.4\,GHz data to
derive the peak position of the source. The position of
$\alpha$=$08^{\rm h}31^{\rm m}41^{\rm s}.576$$\pm$$0^{\rm s}.006$,
$\delta$=$+52^\circ45'14''.66$$\pm$$0''.03$ is in good agreement with
the peak of the 4.5\,GHz continuum (source designation from now on:
VLA\,J083141+524514).  From the unconvolved, naturally weighted
emission line map, we derive a peak flux density of 83 $\pm$
10\,$\mu$Jy beam$^{-1}$. Note that VLA\,J083141+524514 is detected in
none of the other bands, 3$\sigma$ upper limits are given in
Tab.~\ref{t3}.

We also imaged APM\,08279+5255 with uniform weighting (Fig.~\ref{f9}).
The emission is clearly resolved into two peaks at this higher
resolution, which coincide with the \aco\ peaks as indicated by the
plus signs.  Interestingly, both peaks appear to have the same
brightness at 8.4\,GHz (205 $\pm$ 20\,$\mu$Jy beam$^{-1}$ for the
north-eastern peak, and 204 $\pm$ 20\,$\mu$Jy beam$^{-1}$ for the
south-western peak; corresponding to rest-frame peak brightness
temperatures of 126$\pm$12 and 125$\pm$12\,K). This is consistent with
the naturally weighted image where the emission peaks centrally
between the \aco\ peak positions (as both components are blended due
to lower resolution).  This different image brightness ratio relative
to CO may be due to two non-thermal radio continuum components within
the source with different spectral indices that are lensed
differentially. VLA\,J083141+524514 remains unresolved at this higher
resolution.

\begin{figure}
\epsscale{1.22}
\plotone{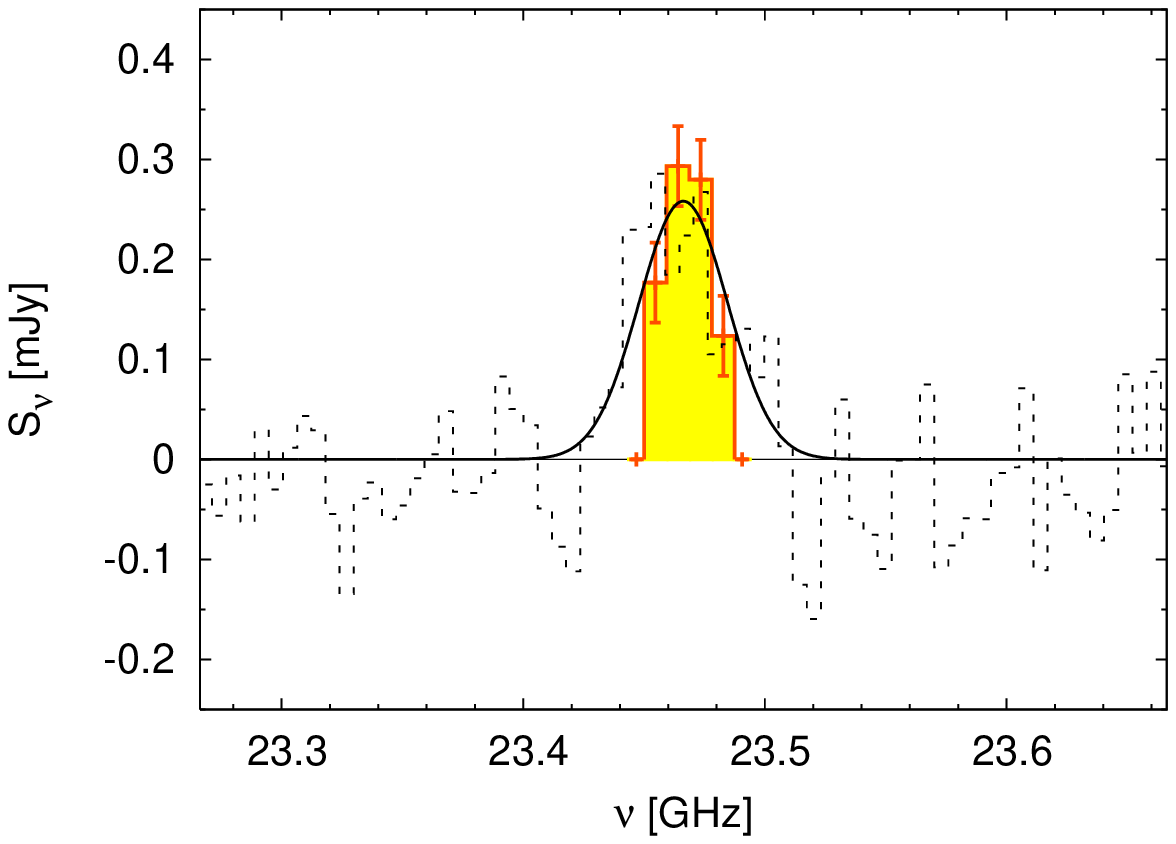}
\caption{
GBT spectrum of \aco\ emission from APM\,08279+5255 (Riechers et al.\
\citeyear{rie06}) at a resolution of 5.86\,MHz (75\,\kms, dashed line).
The rms per channel is 65\,$\mu$Jy. The thin black line shows a
Gaussian fit to the data. Overplotted is the peak flux densities of
our new VLA \aco\ data in the four 120\,\kms\ channels shown in
Fig.~\ref{f3}, but tapered to the resolution shown in
Fig.~\ref{f5}. The error bars indicate the rms per channel of
40\,$\mu$Jy. This overlay shows that the GBT (single-dish) and the VLA
(interferometer) detect the same amount of flux within the errors.
\label{f4}}
\end{figure}

\subsection{Continuum Emission at 14.9\,GHz} \label{14.9ghz}

Continuum emission at 14.9\,GHz was searched for towards
APM\,08279+5255.  We tentatively detect emission at the center of the
\aco\ peak positions. This emission appears to be resolved and
elongated along the same axis as the \aco\ emission.  The peak flux
density measured at this position is 161 $\pm$ 44\,$\mu$Jy
beam$^{-1}$.  However, since the detected flux density is
$<$4$\sigma$, this measurement has to be treated with caution, and
more sensitive data are required to confirm this result.  Convolving
this image with a circular beam of 1.2$''$ diameter as above, we
derive a flux density of 303$\pm$93\,$\mu$Jy beam$^{-1}$, which we
will treat as a tentative detection in the following.  We summarize
all our results in Tab.~\ref{t2}.

\section{Analysis}

\subsection{Spectral Energy Distribution} \label{sed}

Detailed continuum observations are now available for APM\,08279+5255.
Figure \ref{f4a} shows the near-infrared-to-radio
($\sim$1\,$\mu$m--1\,m, or 300\,THz--300\,MHz) spectral energy
distribution (SED).  The mid-IR-to-far-IR part of the spectrum has
been modeled previously (Rowan-Robinson \citeyear{rr2000}; Beelen et
al.\ \citeyear{bee06}), indicating that a two component fit to the
dust emission may be necessary to reproduce the data (a `warm' AGN
dust torus component plus a `cold' starburst component; see also the
multiple-component models by Blain et al.\ \citeyear{bla03}).
Recently, Wei\ss\ et al.\ (\citeyear{wei07}) have presented a model
for the dust continuum that also assumes two components, but which is
additionally constrained by model parameters which result from a study
of multiple rotational transitions of CO and HCN.

However, no detailed analysis of the radio continuum properties of
\apm\ has been conducted so far. We here present two modeling
approaches. The first model aims at finding an overall good fit to the
data while minimizing the number of free parameters. The second model
aims at constraining which fraction of the total radio continuum
emission is due to star formation (i.e., not
AGN-related\footnote{Note that our models do not account for possible
  AGN variability in this source (for which no evidence has been found
  to date).}).

\begin{figure}
\epsscale{1.22}
\plotone{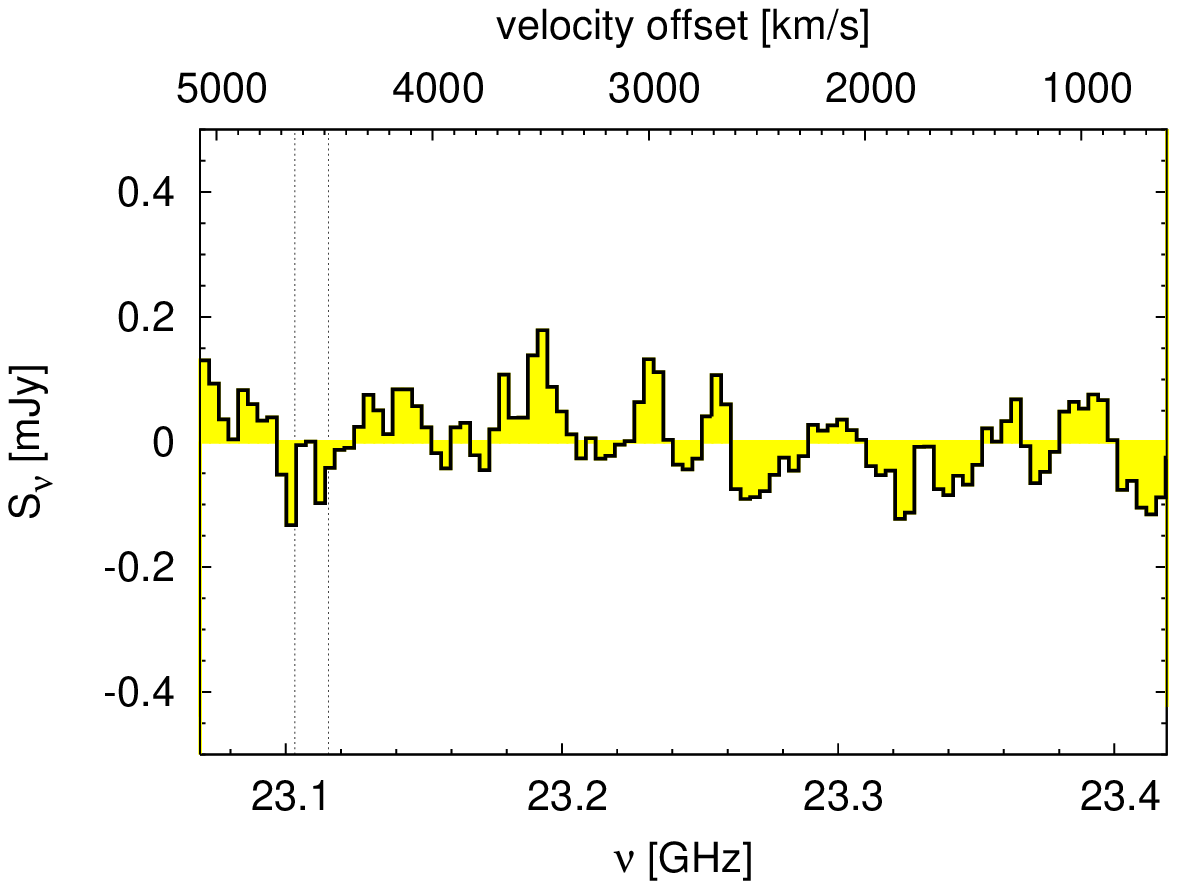}
\caption{Redshifted \acn\ spectrum of APM\,08279+5255 at a resolution
of 3.5\,MHz (45\,\kms), obtained with the GBT. The velocity scale is
relative to the redshifted \aco\ peak ($z$=3.911). The global rms is
65\,$\mu$Jy.  No emission is detected. The dashed lines indicate the
range where the strongest fine structure components of the
CN($N$=1$\to$0, $v$=0, $J$=3/2$\to$1/2) multiplet, i.e.\ the
$F$=3/2$\to$1/2, $F$=5/2$\to$3/2, $F$=1/2$\to$1/2, and $F$=3/2$\to$3/2
transitions, are expected to peak at $z$=3.911. If the CN emission
comes from a similar region than the CO emission, these components are
expected to be blended.
\label{f14}}
\end{figure}

\subsubsection{Best Fit SED Model}

As our focus lies on fitting the radio data, we restrict our analysis
to the mid-IR-to-radio part of the spectral data (60\,$\mu$m to
90\,cm). For comparison only, we also show upscaled versions of quasar
templates by Elvis et al.\ (\citeyear{elv94}) and Richards et al.\
(\citeyear{ric06}) on the optical to mid-IR part of the spectrum,
which however are not particularly good fits to \apm's SED.

Thermal dust emission is mainly caused by ionizing and non-ionizing
ultra-violet (UV) radiation from massive stars and (in quasars)
possibly the central AGN. This radiation is processed by the dust and
then re-radiated in the far-infrared wavelength regime. To minimize
the number of free parameters, we adopt the results of Wei\ss\ et al.\
(\citeyear{wei07}) to describe the dust continuum part of the SED. As
discussed by Wei\ss\ et al., this implies dust masses of $M_{\rm
d,cold} = 2.6 \times 10^9\,\mu_{L}^{-1}\,$M$_{\odot}$ and $M_{\rm
d,warm} = 7.0 \times 10^7\,\mu_{L}^{-1}\,$M$_{\odot}$ and dust
temperatures of $T_{\rm d,cold}=75$\,K and $T_{\rm d,warm}=220$\,K for
the `cold' and `warm' dust components, a dust disk with a magnified
equivalent radius of $r_0 = 1300$\,pc, relative area filling factors
of 78\% and 22\%, and a dust absorption coefficient exponent of
$\beta$=2 to describe its frequency dependence ($\mu_{L}$ denotes the
lensing magnification, the index d denotes thermal dust emission).
Using these parameters and equations 3, 4, 5, and 6 of Wei\ss\ et al.\
(\citeyear{wei07}) fully describes the model used here for the dust
emission. The observed flux density of the dust emission will be
denoted as $S_{\nu, {\rm d}}$ in the following. To extend this model
to the radio continuum, we added a double power-law component, which
we fit by the $\chi^2$ minimization technique.  The full expression
for the model flux density takes the form:

\begin{equation}
  S_\nu(\nu)= S_{\nu, {\rm d}}(\nu) + \sum_{i={\rm flat,steep}}^{} c_i \times \nu^{\alpha_i},
\end{equation}

where $\nu$ indicates the frequency in GHz, $c$ indicates a constant
coefficient, $\alpha$ denotes a spectral index, and $i$ indicates the
index denoting the `flat' and `steep' power-law components.

To solve equation (1), we attempted to fit the observed SED with three
degrees of complexity. In the first step, we fixed the dust term to
the model parameters by Wei\ss\ et al.\ (\citeyear{wei07}, see above),
and fitted the `flat' and `steep' components with a four-parameter
double power law least-squares fit. The best fits to the `flat'
component result in $\alpha_{\rm flat}$ values close to -0.1. However,
the fitting uncertainty of this component is quite large due to the
relatively small region where this component actually dominates the
emission.  In the second step, we thus fixed $\alpha_{\rm flat}$ to
-0.1.  The remaining three free parameters are fairly well
constrained. Our best-fitting model to the composite spectrum then
gives a `steep' power law index of $\alpha_{\rm steep}=-1.31 \pm
0.27$, and `steep' and `flat' coefficients of $c_{\rm steep}=1.09 \pm
0.13$ and $c_{\rm flat}=0.47 \pm 0.06$. The `steep' power law index
suggests that APM\,08279+5255 may be an ultra-steep spectrum source
($\alpha_{\rm steep}$$<-1.2$; e.g., Bondi et al.\ \citeyear{bon07}),
which may indicate that it evolves in a dense environment (Klamer
\etal\ \citeyear{kla06}).  In the third step, we left single or pairs
of parameters of the Wei\ss\ et al.\ (\citeyear{wei07}) dust model as
free parameters to test the stability of the solution. This however
did not further improve our best solution. Our final SED model is
shown in the top panel of Fig.\ \ref{f4a}.

\begin{deluxetable}{lcc}
\tabletypesize{\scriptsize}
\tablecaption{Continuum Fluxes of VLA\,J083141+524514.\label{t3}}
\tablehead{
Frequency & $S_{\nu}$ & Configuration\\
& [$\mu$Jy] & }
\startdata
1.4\,GHz  & $<$100 & A\\
4.5\,GHz  & 180$\pm$40 & B \\
8.4\,GHz  & 83$\pm$10 & BnA/B\\
14.9\,GHz & $<$130 & B\\
23.5\,GHz & $\leq$40 & BCD\\
46.9\,GHz & $<$200 & CD \\
\vspace{-3mm}
\enddata
\tablecomments{
Source is detected at position $\alpha=08^{\rm h}31^{\rm m}41^{\rm s}.576$, $\delta=+52^\circ45'14''.66$.
}
\end{deluxetable}

\subsubsection{Model SED of a Star-forming Galaxy}

\begin{figure*}
\epsscale{0.97}
\plotone{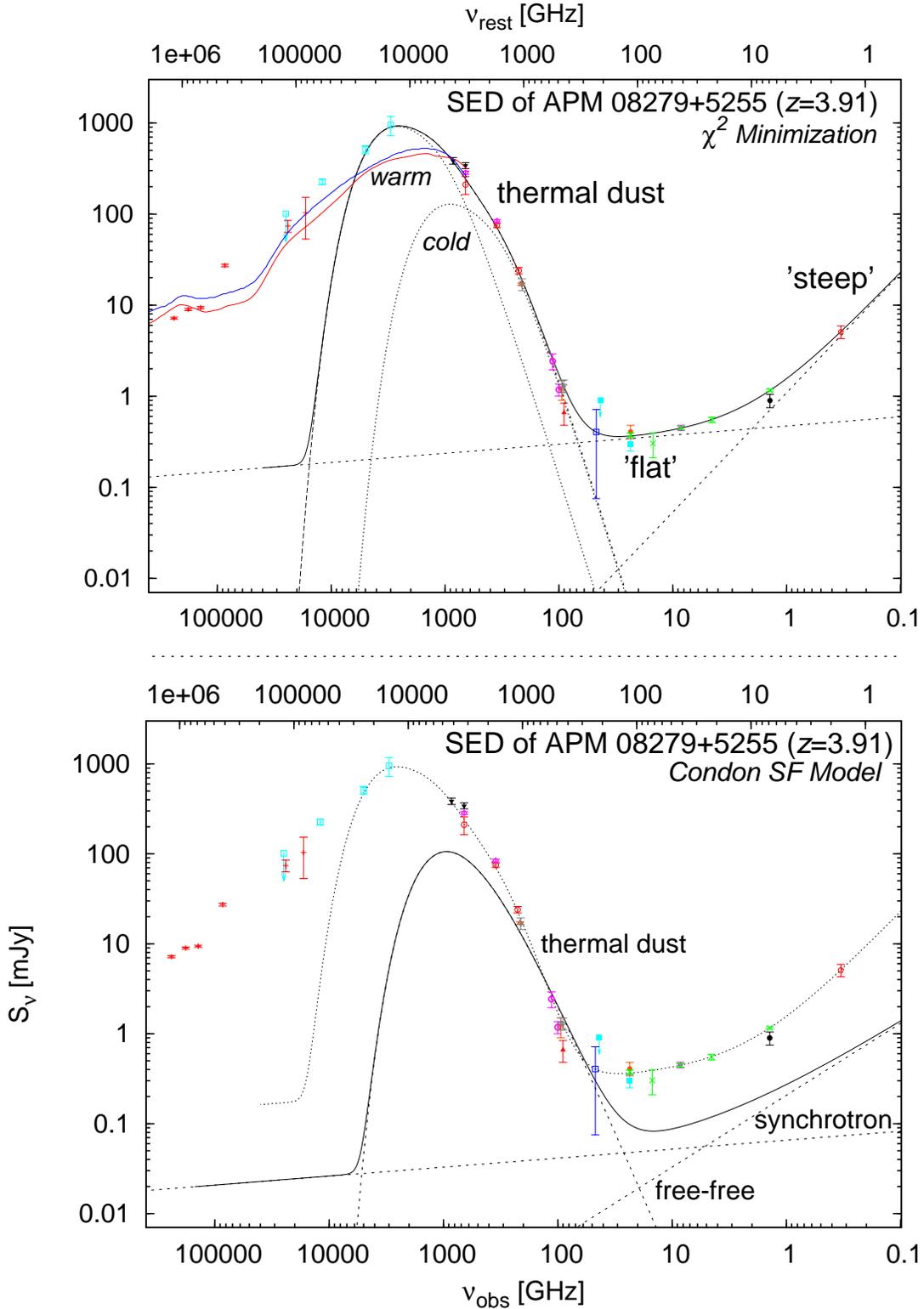}
\vspace{-4mm}

\caption{ 
Near-infrared-to-radio spectral energy distribution (SED) of
APM\,08279+5255. The crosses indicate the measurements described in
this paper, and the open square gives an estimate for the 46.9\,GHz
continuum flux. All other signs indicate detections and upper limits
from the literature (White et al.\ \citeyear{whi97}; Irwin et al.\
\citeyear{irw98}; Lewis et al.\ \citeyear{lew98}; Ibata et al.\
\citeyear{iba99}; Downes et al.\ \citeyear{dow99}; Egami et al.\
\citeyear{e2000}; Papadopoulos et al.\ \citeyear{pap01}; Barvainis \&
Ivison \citeyear{bi02}; Lewis et al.\ \citeyear{lew02}; Wagg \etal\
\citeyear{wag05}, \citeyear{wag06}; Beelen et al.\ \citeyear{bee06};
Ivison \citeyear{ivi06}; and Wei\ss\ et al.\ \citeyear{wei07}).  {\em
Top}: The solid line represents a model fit to the thermal dust
continuum, and a two-component power-law fit to the radio continuum
emission.  The components are indicated by the dashed lines. The
parameters for the model for the dust part of the continuum are
adopted from Wei\ss\ et al.\ (\citeyear{wei07}); the dotted lines
indicate the `cold' and `warm' subcomponents of their model.  The
parameters of the `flat' and `steep' spectrum components are
determined by least-squares power law fitting. For comparison,
upscaled versions of the `radio-quiet' quasar template (red) from
Elvis et al.\ (\citeyear{elv94}) and the `all' quasar template (blue)
from Richards et al.\ (\citeyear{ric06}) are overplotted on the
high-$\nu$ part of the SED.  {\em Bottom}: The solid line represents a
model of the thermal dust, free-free, and synchrotron emission SED
components of a star-forming galaxy, as described by Condon
(\citeyear{con92}). The thick dashed lines indicate those
components. Only the `cold' dust subcomponent which is dominated by
heating from star formation is considered for the model. The thin
dashed line shows the model in the top panel for comparison (see text
for more details).
\label{f4a}}
\end{figure*}

Due to the fact that \apm\ is a composite AGN-starburst system, it is
likely that both components contribute to the continuum emission at
radio wavelengths.  Note that the `warm' dust component in the Wei\ss\
\etal\ model of \apm, which may be dominated by AGN heating,
contributes $\sim$90\% to the far-infrared (FIR) luminosity. This
finding implies that a clear separation of AGN and starburst
contributions to the radio luminosity may be possible.

We here investigate what fraction of the radio emission can be
produced by the starburst component in \apm. Based on the SED of the
Milky Way and first principles, Condon (\citeyear{con92}) has
developed a model for the radio-through-FIR emission from `normal'
galaxies, which is solely attributed to a stellar origin. This model
contains three components: one for thermal dust emission, one for
thermal (mostly free-free) radio emission, and one for non-thermal
(mostly synchrotron) radio emission. The two radio components are
described by power laws. This model also fits the starburst galaxy M82
very well.

Free-free emission is either associated with the ionized gas in
H{\scriptsize II} regions surrounding hot young stars, or the
environment of the active galactic nucleus, and therefore an indicator
for the total photoionization rate. If the thermal, optically thin
free-free (or thermal bremsstrahlung) radiation emerges from a
10$^4$\,K photoionized gas, a spectral index of $\alpha_{\rm ff}=-0.1$
(notation analogue to above) is expected, and thus assumed in the
Condon model.  The free-free power-law coefficient $c_{\rm ff}$ then
parameterizes the emission measure, the source size and the electron
temperature.  However, note that free-free emission can only be
described by a simple power law if the emission is optically thin.

The (non-thermal) synchrotron emission is optically thin at typical
radio wavelengths, and produced in supernova explosions and remnants
(SNRs) through acceleration of cosmic ray electrons, spiraling in a
galaxy's magnetic field.  Detection of strong synchrotron emission
from relativistic electrons thus indicates the presence of a
large-scale magnetic field (Turner et al.\ \citeyear{tur98}). In
starburst galaxies, it may indicate a concentration of SNRs associated
with localized and intense star formation.  Synchrotron emission
usually shows a spectral index of $-1.2 \leq \alpha_{\rm sync} \leq
-0.4$, with a typical value of --0.8, which is assumed in the Condon
model.  Here, the synchrotron power-law coefficient $c_{\rm sync}$
parameterizes the type II supernova rate (assuming that supernova
remnants dominate the non-thermal synchrotron emission).

In the Condon (\citeyear{con92}) model, all three components (dust
heating, free-free, and synchrotron) are directly proportional to the
star-formation rate (SFR) in a galaxy.  The dust emissivity $\beta$,
the dust temperature $T_{\rm d}$, and the SFR are the only model
parameters. We here adopt $\beta$=1.5 from Condon's model.  Due to the
fact that the `warm' dust component in \apm\ may be dominantly heated
by the AGN, we here only consider the `cold' component, which likely
traces the dominant fraction of the dust heated by star formation. We
thus adopt $T_{\rm d}$=75\,K. The only remaining free parameter thus
is the SFR. Assuming a SFR of
4000\,$\mu_{L}^{-1}$\,M$_\odot$\,yr$^{-1}$ provides a reasonable fit
to the spectrum of the cold dust of \apm\ (see Fig.\ \ref{f4a},
bottom).  However, this model only recovers $\sim$20\% of the radio
emission.  Yun \& Carilli (\citeyear{yun02}) suggested some changes to
the Condon model\footnote{They suggest a different initial mass
function, an update to the H$\alpha$ normalization, and a small
correction factor for the supernova rate.}, which would result in
about a factor of 2 less radio flux compared to the original Condon
model. As a next step, we investigated if including the $T_{\rm
d}$=220\,K component (under the assumption that it is also heated by
star formation) improves the fit. In this model, the peak of the dust
spectrum limits the SFR from the `warm' component to
$\sim$1500\,$\mu_{L}^{-1}$\,M$_\odot$\,yr$^{-1}$. Assuming such high
values however requires to reduce the SFR of the `cold' component to
only few hundreds $\mu_{L}^{-1}$M$_\odot$\,yr$^{-1}$ in order to still
reproduce the low-$\nu$ end of the dust bump. This significantly
decreases the model-predicted radio flux, and thus gives an overall
worse fit. The full radio flux could be recovered by including a very
cold, $\sim$10-15\,K dust component that is maintained by an extreme
15,000-20,000\,$\mu_{L}^{-1}$\,M$_\odot$\,yr$^{-1}$ starburst. Such a
component could exist in theory, but would require to be spread out
over large scales in order to exhibit such low dust temperatures, and
thus is basically ruled out by high spatial resolution studies of this
source. Although it is difficult to assess the validity of a simple
Condon model for a warm, starbursting quasar like \apm, this result
leads us to conclude that the radio continuum emission in this source
may be dominanted by the AGN.  This would be in agreement with our
best-fit model, as it implies an ultra-steep radio spectrum component,
which is usually associated with a source where AGN-related emission
dominates the energy output at radio frequencies (e.g., high redshift
radio galaxies and quasars; Athreya
\etal\ \citeyear{ath97}; Pentericci \etal\ \citeyear{p2000}).  The
ultra-steep spectrum component in AGN-driven radio sources is often
associated with extended radio emission.  Interestingly, the best-fit
model also requires a relatively bright `flat'-spectrum component to
originate from the radio AGN.  The core of the archetypical radio
galaxy Cygnus A has a flat spectrum at radio wavelengths, which
dominates over the steep-spectrum emission from the hotspots at higher
frequencies (e.g., Eales \etal\ \citeyear{eal89}). A similar
`composite' effect (extended steep-spectrum component, flat core
component) may cause the apparent flattening of the radio spectrum of
\apm.

\subsection{Comparison to Previous CO(1--0) Imaging} \label{history}

The \aco\ transition line in APM\,08279+5255 was targeted twice before
with the VLA, first at C array resolution (1.5\,$''$$\times$1.4\,$''$,
Papadopoulos et al.\ \citeyear{pap01}), and then followed up at higher
B array resolution (0.39\,$''$$\times$0.28\,$''$, Lewis et al.\
\citeyear{lew02}). Both observations were setup in quasi-continuum
mode, i.e.\ lacking any spectral information of the \aco\ line. In
both observations, \aco\ line and 23.3649\,GHz continuum emission were
detected towards the source. Papadopoulos et al.\ (\citeyear{pap01})
detect unresolved continuum emission at a flux density level of 0.30
$\pm$ 0.05\,mJy. Lewis et al.\ (\citeyear{lew02}) detect continuum
emission at a flux density of 0.41 $\pm$ 0.07\,mJy, and the resolved
structure coincides with our continuum map in structure and position.
Both flux densities are in agreement with our result within the
errors, although some intrinsic variability cannot be excluded.

Papadopoulos et al.\ (\citeyear{pap01}) report the detection of
resolved \aco\ emission towards APM\,08279+5255, which extends over
scales of $\gtrsim$7\,$''$$\times$2.25\,$''$ (their 3$\sigma$ contour)
at a resolution convolved to 2.25\,$''$$\times$2.25\,$''$ (7\,$''$
correspond to $\sim$30\,kpc at the target redshift of 3.911). They
quote an integrated flux for the `nuclear \aco\ emission' (the inner
$\sim$1$''$) of 0.150 $\pm$ 0.045\,Jy \kms.

From our new VLA observations, we derive an integrated \aco\ flux of
0.168 $\pm$ 0.015\,Jy \kms\ assuming a line FWHM of 556 $\pm$
55\,\kms\ (Riechers et al.\ \citeyear{rie06}; width of our bandpass:
558\,\kms). This corresponds to a \aco\ line luminosity of $L'_{\rm
  CO(1-0)}$ = (10.6$\pm$0.9)$\times$10$^{10}\,\mu_{L}^{-1}$\,K\,\kms\,pc$^2$
(see Tab.~\ref{t4}).  From the \aco\ spectrum obtained with the GBT,
Riechers et al.\ (\citeyear{rie06}) find 0.152 $\pm$ 0.020\,Jy \kms.
The 3$\sigma$ contour of the extended \aco\ reservoir of Papadopoulos
et al.\ (\citeyear{pap01}, their Fig.~1) corresponds to a flux density
of 120\,$\mu$Jy beam$^{-1}$. In our \aco\ map which was tapered to a
linear resolution of $\sim$1.2$''$ (Fig.\ \ref{f5}, similar to the
resolution of their observations), this would correspond to
6.3$\sigma$. Even at less than half this flux level
(3$\sigma$=57\,$\mu$Jy beam$^{-1}$), we find no evidence for any
extended flux.  We note that our observations even include shorter (D
array) baselines than Papadopoulos et al.\ (\citeyear{pap01}) and are
therefore more sensitive to extended structure.  We thus exclude the
possibility that our observations outresolve the extended emission,
and conclude that there is no bright, extended \aco\ reservoir in
\apm.

Lewis et al.\ (\citeyear{lew02}) find resolved \aco\ emission on
sub-arcsec scale at moderate signal-to-noise. The peak of the emission
reported in their paper roughly coincides with our north-eastern \aco\
peak. However, our new, more sensitive (by a factor of 2 in terms of
rms noise) high-resolution maps do not show the extended structure
that is present in their \aco\ map, which we thus conclude to be a
noise artifact.  Lewis et al.\ (\citeyear{lew02}) derive an integrated
\aco\ flux of 0.22 $\pm$ 0.05\,Jy \kms\ from their observations, in
agreement with the three independent results given above within the
errors (adopting only the `nuclear \aco\ emission' from the report of
Papadopoulos et al.).

\begin{deluxetable*}{lcccc}
\tabletypesize{\scriptsize}
\tablecaption{Line Intensities, Luminosities, and Ratios.\label{t4}}
\tablehead{
Line & $I$ & $L'$ & $L'/L'_{\rm CO(1-0)}$ & data reference\\
& [Jy\,\kms ] & [10$^{10}$\,$\mu_{L}^{-1}$\,K\,\kms\,pc$^2$] & &}
\startdata
\aco &  0.168 $\pm$ 0.015 & 10.6 $\pm$ 0.9 & 1 & 1,2 \\
\bco &  0.81  $\pm$ 0.18  & 12.8 $\pm$ 2.9 & 1.21 $\pm$ 0.30 & 1,3 \\
\dco &  3.7   $\pm$ 0.2   & 14.7 $\pm$ 0.9 & 1.39 $\pm$ 0.16 & 4 \\
\fco &  6.7   $\pm$ 1.2   & 11.8 $\pm$ 2.1 & 1.12 $\pm$ 0.23 & 4 \\
\ico & 11.8   $\pm$ 0.6   &  9.2 $\pm$ 0.4 & 0.89 $\pm$ 0.10 & 4 \\
\jco & 11.9   $\pm$ 2.0   &  7.5 $\pm$ 1.2 & 0.71 $\pm$ 0.14 & 4 \\
\kco & 11.3   $\pm$ 1.9   &  5.9 $\pm$ 1.0 & 0.56 $\pm$ 0.11 & 4 \\
\acn &           $<$0.06  &         $<$3.9 &         $<$0.37 & 1 \\
\vspace{-3mm}
\enddata
\tablerefs{[1] this work; [2] Riechers et al.\ \citeyear{rie06};
[3] Papadopoulos et al.\ \citeyear{pap01}; [4] Wei\ss\ et al.\ \citeyear{wei07}.
}
\tablecomments{${}$Luminosities are apparent values uncorrected for
gravitational magnification. They are derived as described by Solomon
\etal\ (\citeyear{sol92}): $L'[{\rm K \ts km\ts s^{-1} pc^2}] = 3.25
\times 10^7 I \nu_{\rm obs}^{-2} D_{\rm L}^2 (1+z)^{-3} \mu_{L}^{-1}$, 
where $I$ is the velocity-integrated line flux in Jy \kms,
$D_{\rm L}$ is the luminosity distance in Mpc, $\nu_{\rm obs}$ is the
observed frequency in GHz, and $\mu_{L}$ is the lensing
magnification factor.  
}
\end{deluxetable*}

\subsection{On the 46.9\,GHz Continuum Emission} \label{bco-cont}

As mentioned previously, the continuum emission at the line frequency
(46.9\,GHz) was not monitored in parallel during the (archival) \bco\
observations. Papadopoulos et al.\ (\citeyear{pap01}) report a limit
of $3\sigma=0.9$\,mJy on the continuum emission at 43.3\,GHz. This
corresponds to $\sim$55\% of the \bco\ peak flux. Therefore, the
`real' \bco\ flux towards APM\,08279+5255 may be lower by up to a
factor of 2. Based on our SED modeling, we now attempt to set
additional limits to the continuum contribution to the emission
detected at the \bco\ frequency to attain some tighter constraints.

A first-order estimate for the 46.9\,GHz continuum flux is provided by
examining the overall spectral energy distribution of APM\,08279+5255.
The SED model presented in Fig.\ \ref{f4a} predicts a continuum flux
at 46.9\,GHz of 405\,$\mu$Jy. This estimate is plotted as the square
symbol in Fig.\ \ref{f4a}.  A lower constraint is provided by the fit
to the thermal part of the SED (75\,$\mu$Jy), and an upper constraint
is given by linear interpolation of the 23\,GHz and 94\,GHz continuum
flux measurements (715\,$\mu$Jy). For illustration, these constraints
are plotted as error bars to this model-predicted data point. We
however note that the 23\,GHz and 94\,GHz measurements were obtained
at different epochs, therefore this estimate does not account for
possible variability of the millimeter continuum.

As another estimate, we assume that both the emission from \aco\ and
\bco\ are fully thermalized and optically thick. The \bco\ peak flux
should then be four times higher than the \aco\ peak flux. For the
\aco\ peak flux, an average over the whole 558\,\kms\ bandpass
(210$\pm$19\,$\mu$Jy) is assumed to match the velocity resolution of
the \bco\ observations as closely as possible. In comparison to the
measured flux at the \bco\ frequency (1637$\pm$65\,$\mu$Jy), this
predicts a contribution of the continuum flux of $\sim$800\,$\mu$Jy,
which is twice as high as the SED-based estimate.  This discrepancy
may be explained by different optical depths of the emission from
different rotational CO transitions. If the opacity of the \aco\
emission is much lower than that of the higher $J$ transitions, it may
actually have a lower luminosity than the emission from these
transitions [contrary to the effects of (sub-)thermal excitation].
This would allow the \bco\ line peak flux to be more than four times
higher than the \aco\ peak flux. Note that such an effect may actually
be observed towards \apm\ in the $J$$\geq$4 transitions, as the
\dco/\aco\ luminosity ratio is $>$1 (Wei\ss\ et al.\ \citeyear{wei07};
see also Tab.~\ref{t4}).

We conclude that APM\,08279+5255 likely exhibits a 46.9\,GHz continuum
flux of at least 0.4\,mJy, i.e.\ 25\% of the measured \bco\ flux. We
thus estimate the `real' \bco\ line peak flux to be $\sim$1.2\,mJy.
From this, we derive a \bco\ line luminosity of $L'_{\rm
  CO(2-1)}$=(12.8$\pm$2.9)$\times$10$^{10}\,\mu_{L}^{-1}$\,K\,\kms\,pc$^2$.

\subsection{A Limit for the CN Luminosity} \label{acn-model}

\begin{figure}
\epsscale{1.22}
\plotone{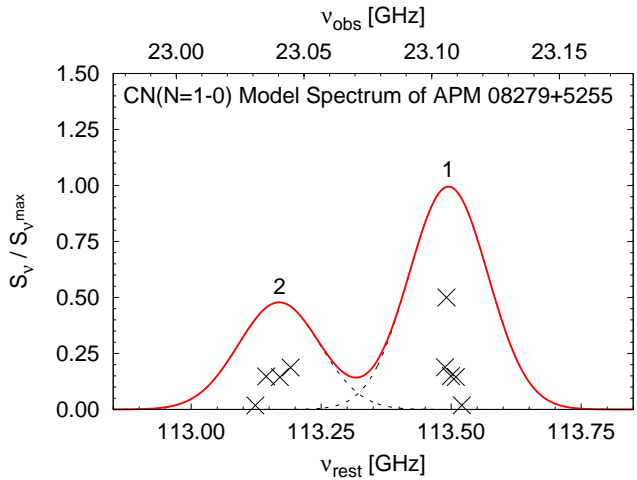}
\caption{Model spectrum of the \acn\ emission in APM\,08279+5255. The
  horizontal axes indicate the rest and observed frequencies.  The
  vertical axis indicates the predicted intensity, normalized to the
  peak of the emission.  The relative intensities of the fs and hfs
  components are computed for the LTE case. The crosses indicate the
  hfs components.  The dashed lines are Gaussian fits to the summed
  intensities in the two main frequency bins, assuming the average
  width of the higher $J$ CO lines (Wei\ss\ et al.\ \citeyear{wei07})
  for the subcomponents (labeled '1' and '2'). The solid line is a sum
  of all components, and indicates the model-predicted \acn\ line
  shape. \label{f18}}
\end{figure}

Due to its fine structure (fs) and hyperfine structure (hfs)
splitting, the \acn\ emission line is distributed over 9
lines\footnote{We assume that CN is in its ground electronic state
  (${}^2\Sigma$), and that the spins couple according to Hund's case
  (b) coupling scheme: $\vec{N} + \vec{S} = \vec{J}$, (fs coupling)
  and $\vec{J} + \vec{I_{\rm C}} + \vec{I_{\rm N}} = \vec{F}$ (hfs
  coupling). Here, $\vec{N}$ is the rotational angular momentum
  vector, $\vec{S}$ is the electronic spin, and $\vec{I}$ is a nuclear
  spin.} in its $v$=0 vibrational state. These lines are distributed
in 2 main components separated by about 350\,MHz in the rest frame.
While the hfs transitions within these main components are blended if
their kinematical broadening is similar to that of the CO/HCN lines in
APM\,08279+5255, the components themselves are still clearly
separated. To obtain a proper limit on the integrated \acn\ line flux
(and thus on the total CN luminosity), we calculated a synthetic line
profile, assuming optically thin emission in the Local Thermodynamic
Equilibrium (LTE) case to derive the relative intensities of the hfs
components. The relative intensities of the components were computed
using the laboratory data from Skatrud et al.\ (\citeyear{ska83}), and
approximation (4) of equation (1) of Pickett et al.\
(\citeyear{pic98}).  Assuming that all \acn\ hfs components are
kinematically broadened in the same way as the higher $J$ CO lines
($\sim$500\,\kms\ FWHM; see Wei\ss\ et al. \citeyear{wei07}), we
obtain the synthetic line profile displayed in Fig.~\ref{f18} (solid
line).  The two main components (dashed lines) have a peak strength
ratio of 2:1. The brightest main component is fully covered by our
bandpass, the second component, however, was close to the edge, and
thus only covered in part by the part of the spectrum that was
properly calibratable. For the derivation of the limit on the
integrated \acn\ line flux, we thus consider the peak flux limit for
the brightest component to derive the relative intensity limits for
all subcomponents in the LTE case.  Under these assumptions, we derive
a limit of $I_{\rm CN}$$<$0.06\,Jy\,\kms\ for the integrated line
flux, and thus a limit of $L'_{\rm
  CN}$$<$3.9$\times$10$^{10}\,\mu_{L}^{-1}$\,K\,\kms\,pc$^2$ for the
line luminosity. This corresponds to a luminosity ratio limit of
$L'_{\rm CN}$/$L'_{\rm CO}$$\leq$0.37.  Note that Gu\'elin et al.\
(\citeyear{gue07}) recently observed the \dcn\ transition towards
\apm; however, due to strong blending with the \ehnc\ line, it remains
unclear whether or not the line is detected.

\subsection{Radio Luminosity} \label{radio}

In general, the monochromatic rest-frame 1.4\,GHz radio luminosity is
defined as:

\begin{equation}
L_{\rm 1.4\,GHz} = \frac{4 \pi D_{\rm L}^2}{(1+z)^{1+\alpha_{1.4}}}S_{\rm 1.4\,GHz} .
\end{equation}

However, we here calculate $L_{\rm 1.4\,GHz}$ based on the
(rest-frame) 1.4\,GHz flux predicted by our best-fit SED model. In the
above equation, this would correspond to using a spectral index of
$\alpha_{1.4}$=--1.05.  We thus obtain $L_{\rm
1.4\,GHz}$=(1.90$\pm$0.05)$\times$10$^{26}$\,$\mu_{L}^{-1}$\,W\,Hz$^{-1}$
(not lensing corrected), about 10 times the radio luminosity of M87
($L_{\rm 1.4\,GHz}$=(1.76$\pm$0.07)$\times$10$^{25}$\,W\,Hz$^{-1}$;
Laing \& Peacock \citeyear{lp80}). In an analogous manner, we derive a
5\,GHz radio luminosity of $L_{\rm
5\,GHz}$=(4.75$\pm$0.34)$\times$10$^{25}$\,$\mu_{L}^{-1}$\,W\,Hz$^{-1}$.
Here, the model prediction would correspond to a spectral index of
only $\alpha_{5}$=--0.67.

Studies of the radio luminosity function of quasars and its evolution
have identified a bimodality in the source distribution, separating
them into radio-quiet and radio-loud quasars.  However, two generally
different definitions of radio loudness have been put forward, one
based on the optical-to-radio flux ratio (e.g., Schmidt
\citeyear{sch70}; Kellerman \etal\ \citeyear{kel89}), and one only
based on the monochromatic radio luminosity (e.g., Peacock \etal\
\citeyear{pea86}; Schneider \etal\ \citeyear{sch92}).  However,
Ivezi\'c \etal\ (\citeyear{ive02}) argue that for optically selected
quasars, both definitions are similar within the flux-limited samples
that were examined.  We thus restrict our analysis to the radio
luminosity-based estimate here.  Due to different assumptions and
definitions, a range of threshold values for the radio loudness
definition is found in the literature.  For the radio-based
definition, we here adopt the range of $L_{\rm
5\,GHz}$=(1.0--5.6)$\times$10$^{25}$\,W\,Hz$^{-1}$ as given by
Schneider \etal\ (\citeyear{sch92}). Changing the cutoff within this
range does not change the relations they find dramatically. Even
without correcting for gravitational lensing, \apm\ clearly falls
within the transition region between radio-quiet and radio-loud
sources. As this source is known to be substantially gravitationally
magnified (see also discussion below), we conclude that it is a
radio-quiet quasar by this definition.

\subsection{Morphology at X-Ray to Radio Wavelengths} \label{nir}

\begin{figure*}
\vspace{-4mm}

\epsscale{1.15}
\plotone{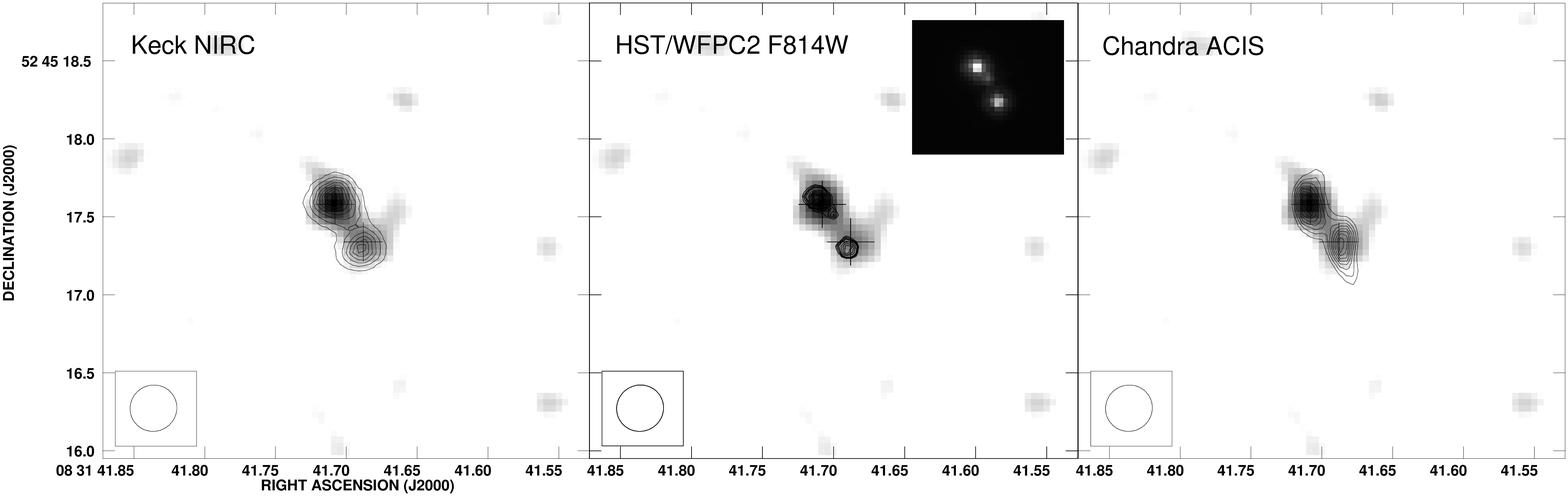}
\vspace{-7.5mm}

\caption{Morphology of APM\,08279+5255 at different wavelengths. {\em
 Left}: Contours of 2.2\,$\mu$m continuum emission (Egami et
 al.~\citeyear{e2000}) overlaid on the \aco\ emission. {\em Middle}:
 Contours of 814\,nm continuum emission (HST archive) overlaid on
 the \aco\ emission. The inset shows the optical image only. {\em
 Right}: Contours of hard X-ray emission (Chartas et
 al.~\citeyear{cha02}) overlaid on the \aco\ emission. Contours in
 all panels are 20\%, 30\%, ..., 90\% of the peak flux of the
 emission, but with 23.3\% and 26.7\% added in the middle panel to
 emphasize the third image. The CO greyscale scales between
 2$\sigma$ and the peak flux of the emission (see
 Fig.~\ref{f1}). The crosses indicate the CO peak positions. The
 size of the synthesized beam of the CO observations is indicated
 in the bottom left corner.\label{f1a}}
\end{figure*}

Optical imaging of APM\,08279+5255 with the Hubble Space Telescope
(HST) Near Infrared Camera and Multi-Object Spectrometer (NICMOS) has
revealed three pointlike, almost collinear images with very similar
colors (Ibata et al.~\citeyear{iba99}). A spectroscopic follow-up
study with the Space Telescope Imaging Spectrograph (STIS)
spectrograph on board HST has shown that these three images have
essentially identical spectral shapes, and thus are, indeed, all
images caused by gravitational lensing (Lewis et
al.~\citeyear{lew02b}). These three optical images A, B, and
C\footnote{By convention, A is the brightest image, and C is the
faintest image.} are separated by $d$(A,B)=0.377$''$$\pm$0.002$''$ and
$d$(A,C)=0.150$''$$\pm$0.006$''$, and have brightness ratios of
$b$(A,B)=0.773 $\pm$ 0.007 and $b$(A,C)=0.175 $\pm$ 0.008, i.e.,
$b$(A+C,B)$\simeq$0.66.  The north-eastern CO peak corresponds to a
blend of optical images A+C, and the south-western CO peak to optical
image B. The CO image peak brightness ratio of $b$(SW,NE)=0.66 $\pm$
0.15 thus agrees very well with the optical observations.  The
separation of the CO images is by $\sim$20\% lower than that of
optical images A and B, which agrees with the assumption that the
north-eastern CO component is actually a blend of optical images A+C,
which we will assume in the following.

In their analysis, Ibata et al.\ (\citeyear{iba99}) find an offset of
0.6$''$ between the positions of HST NICMOS image A and the 8.4\,GHz
continuum source, which is not thought to be real. From our analysis,
we find that the observations in six different radio frequency bands
are consistent in position among each other to high precision (and
with the radio positions from previous studies). We also find that the
offset between the radio position and a more recent image taken with
HST Wide-Field Planetary Camera (WFPC2; HST archive, see also
Fig.~\ref{f1a}) is $<$0.2$''$. This implies a large offset between
the WFPC2 and NICMOS images, which are close in wavelength and likely
dominated by compact, parsec-scale AGN emission which is expected to
be spatially coincident. In the following, we thus assume that these
offsets are due to HST astrometric calibration errors, and shift all
images at other wavelengths to the radio coordinates.

The three images of APM\,08279+5255 were also resolved in the
(observed frame) near-infrared (NIR) continuum at 2.2\,$\mu$m in
ground-based adaptive optics observations with the Keck telescopes
(Egami et al.~\citeyear{e2000}) at a resolution where images A and C
are still significantly blended, similar to our \aco\ observations.
The left panel of Fig.~\ref{f1a} shows an overlay of the 2.2\,$\mu$m
continuum emission and the \aco\ emission. The 2.2\,$\mu$m has been
de-rotated (see caption of Fig.~1 of Egami et al.~\citeyear{e2000} for
the rotation angle) and aligned to the radio coordinates.  The
resolution of the NIR observations is by about a factor of 2 higher
than that of the CO observations, and thus peaks close to the position
of image A rather than between A and C. The image separation of the
two peaks thus is $d$(A,B)=0.38$''$, in agreement with the optical
images (Fig.~\ref{f1a}, middle panel). Taking this into account, it is
striking how similar the morphologies of the NIR and CO emission are,
even though they are expected to emerge from largely different scales
(hot dust around the central AGN vs.\ cold molecular gas in an
extended star-forming ring), and thus to be differentially
lensed. This similarity of morphologies even holds for hard X-ray
emission from the central engine (Chartas et al.~\citeyear{cha02}, see
Fig.~\ref{f1a}, right), which is expected to be by many orders of
magnitude more compact than the CO emission. Even the image separation
$d$(A,B)=0.38$''$$\pm$0.01$''$ of the two main peaks (image A and C
are again significantly blended) is the same as in the optical/NIR,
and thus in agreement with the CO observations under the above
assumptions.
 
In addition, the rest-frame 2.6\,mm continuum underlying the \aco\
line shows a different, more extended structure than the CO
emission\footnote{A possible caveat is that the bandwidth of our
observations is too narrow to image the \aco\ linewings. It however is
highly unlikely that this significantly alters this conclusion.}  (see
Fig.~\ref{f1}).  This may be due to a foreground source that emits in
the continuum, but does not contribute to the line emission due to its
different redshift. Another, probably more likely explanation is that
the different structure is due to differential lensing. If the 2.6\,mm
continuum emission were to be due to star formation, it would be
possible that it is dominated by a free-free component that is not
co-spatial with the CO emission, but extended out to kpc scales. On
the other hand, the SED of \apm\ indicates that a major fraction of
emission at this wavelength may also be AGN-related. If correct, this
would suggest that the differentially lensed emission is due to
100\,pc to kpc-scale outflow from the nucleus.  Higher resolution
observations using a very long baseline radio interferometer are
required to further investigate the nature of this phenomenon. Also,
imaging the rest-frame FIR continuum at comparable resolution may shed
more light on this issue. In the following, we will refer to this
emission as `the extended component'. This differentially lensed,
extended component may also contribute significantly to the continuum
emission at 7.3\,mm (8.4\,GHz observed frame), which also shows a
different structure compared to other wavelengths.

The morphological similarity of the strongly lensed quasar
APM\,08279+5255 on largely different scales ($<$0.1\,pc to few
100\,pc) as well as the different structure of the extended continuum
component (probably few 100\,pc to kpc) severely constrains the lens
configuration, and any valid lens model will have to be able to
reproduce this scaling effect.

\section{Gravitational Lensing} \label{gl}

\begin{figure}
\epsscale{1.21}
\plotone{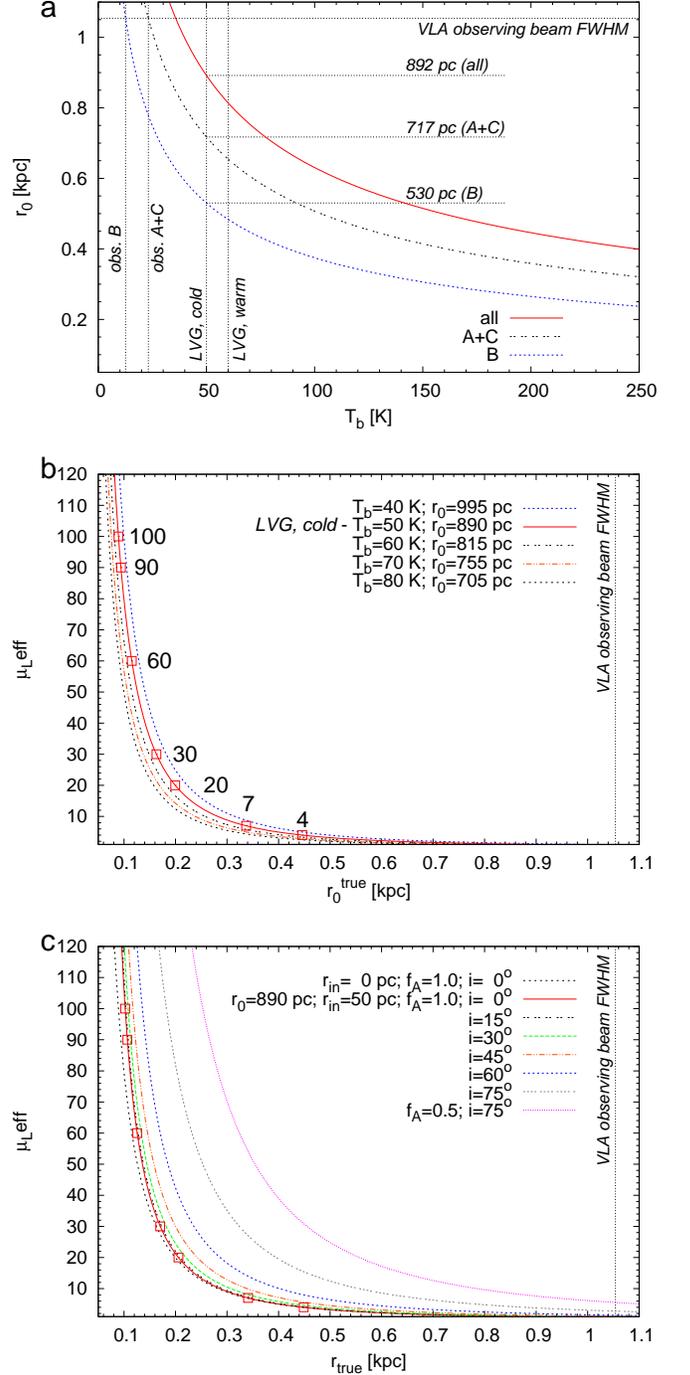}
\vspace{-5.5mm}

\caption{Observational constraints on the intrinsic brightness
  temperature $T_{\rm b}$ and the lensing properties of
  APM\,08279+5255. {\bf a}: Relation between $T_{\rm b}$ and the
  lensed equivalent disk radius $r_0$. The vertical lines indicate the
  observed \aco\ brightness temperatures for the blended images A and
  C and image B, and the LVG predicted brightness temperatures for the
  `cold, dense' (50\,K) and `warm' (60\,K) \aco\ components from the
  model by Wei\ss\ et al.\ (\citeyear{wei07}). The horizontal lines
  indicate the lensed equivalent disk radii for the different images
  for a true $T_{\rm b}$ of 50\,K, and the total $r_0$ for all images.
  The fourth horizontal line indicates the VLA beam radius, and also
  appears in panels {\bf b} and {\bf c}. {\bf b}: Relation between the
  true, unlensed equivalent disk radius $r_0^{\rm true}$ and the
  effective lensing magnification factor $\mu_{L}^{\rm eff}$ for
  different $T_{\rm b}$. For $T_{\rm b}$=50\,K, different
  $\mu_{L}^{\rm eff}$ are highlighted. {\bf c}: Relation between the
  true, unlensed radius $r_{\rm true}$ of the CO ring and
  $\mu_{L}^{\rm eff}$ for the $T_{\rm b}$=50\,K solution assuming an
  inner ring radius $r_{\rm in}$ of 50\,pc for different inclinations
  $i$ and area filling factors $f_{\rm A}$. For comparison, a solution
  without an inner boundary is shown (see text for
  details).\label{f17}}
\end{figure}

\subsection{Direct Constraints on the Lensing Properties from the CO Observations} \label{obslens}

The brightness temperature $T_{\rm b}$ of a lensed source is a
conserved quantity under gravitational lensing, as the latter is a
purely geometrical effect. Due to the fact that APM\,08279+5255 is
resolved in our \aco\ observations, we can derive the observed
brightness temperatures $T_{\rm b}^{{\rm obs},z}$ at redshift $z$ for
images A+C and B directly from the observed peak fluxes using

\begin{equation}
T_{\rm b}^{{\rm obs},z}=\frac{c^2}{2 k_{\rm B} \nu_{\rm obs}^2 \Omega_{\rm beam}}S_\nu ,
\end{equation}

where $\Omega_{\rm beam}$ is the beam solid angle. For Gaussian beam
and source shapes, this equation can be simplified to

\begin{equation}
T_{\rm b}^{{\rm obs},z} \simeq 1.359 \times 10^{-5} \frac{\lambda_{\rm obs}^2}{\theta_{\rm maj}\,\theta_{\rm min}} S_\nu ,
\end{equation}
where $T_{\rm b}$ is given in K, $\lambda_{\rm obs}$ is given in mm,
the beam major and minor axes $\theta_{\rm maj}$ and $\theta_{\rm
min}$ are given in arcsec, and $S_\nu$ is given in $\mu$Jy.  These
brightness temperatures can be converted to rest-frame observed
brightness temperatures $T_{\rm b}^{\rm obs}$ via

\begin{equation}
T_{\rm b}^{\rm obs}=(1+z) T_{\rm b}^{{\rm obs},z} .
\end{equation}

While \aco\ images A+C and B are individually detected in our
observations, their true size is likely smaller than our observing
beam. The observed brightness temperatures $T_{\rm b}^{\rm obs}$ thus
are diluted by the beam (radius: $r_{\rm beam}$) and smaller than the
true $T_{\rm b}$.  The size of the lensed images can be expressed by
an equivalent radius $r_0$, which assumes that each image is a filled,
circular lensed disk. Energy conservation then gives:

\begin{equation}
T_{\rm b}^{\rm obs}/T_{\rm b}=(r_0/r_{\rm beam})^2 .
\end{equation}

This relation is implied in the plot shown in Fig.~\ref{f17}a. The
observed, beam-diluted (Rayleigh-Jeans) brightness temperatures of
images A+C ($T_{\rm b}^{\rm obs, A+C}$ = 23.2\,K) and B ($T_{\rm
  b}^{\rm obs,B}$ = 12.6\,K) are indicated by vertical lines. The
three curves indicate the beam-corrected brightness temperatures for
different equivalent radii (which, by definition, cross with the
observed rest-frame brightness temperatures at $r_0$/$r_{\rm beam}$,
as indicated by the long horizontal line) for images A+C, B, and all
images together.  From Large Velocity Gradient (LVG) modeling of the
\aco\ to \kco\ SLED, Wei\ss\ et al.\ (\citeyear{wei07}) have derived
that the molecular gas in APM\,08279+5255 can be modeled with two gas
components, a `cold, dense' component with a H$_2$ density of $n(\rm
H_2)$=10$^5$\,cm$^{-3}$ and a kinetic gas temperature of $T_{\rm
kin}$=65\,K which contributes about 70\% to the total \aco\
luminosity, and a `warm' component with a H$_2$ density of $n(\rm
H_2)$=10$^4$\,cm$^{-3}$ and a kinetic gas temperature of $T_{\rm
kin}$=220\,K which contributes about 30\% to the total \aco\
luminosity. For the `cold, dense' component, their model predicts a
true \aco\ brightness temperature of $T_{\rm b}^{\rm cold}$=50\,K, and
for the `warm' component, it predicts $T_{\rm b}^{\rm warm}$=60\,K.
Both $T_{\rm b}$ are indicated in Fig.~\ref{f17}a by vertical lines.
Assuming $T_{\rm b}$=50\,K, as predicted by the LVG models for the
dominant gas component, images A+C and B fill equivalent disks with
radii of $r_0^{\rm A+C}$=717\,pc and $r_0^{\rm B}$=530\,pc. This
corresponds to a total lensed equivalent disk with a radius of
$r_0$=892\,pc, consistent with the results of Wei\ss\ et al.\
(\citeyear{wei07}) within the uncertainties. This also implies beam
area filling factors of 46\% and 25\% for images A+C and B. These are
significant fractions of the beam size, and may indicate that
increasing the resolution by only a factor of 2--3 would be sufficient
to detect substructure within the individual images.  Assuming a 30\%
contribution of a warmer component with $T_{\rm b}$=60\,K does not
significantly alter any of these predictions.  Even assuming much
higher $T_{\rm b}^{\rm CO}$ than typically observed in the central
regions of nearby warm, ultra-luminous infrared galaxies (ULIRGs,
e.g., Downes \& Solomon \citeyear{ds98}) of a few hundred Kelvins does
not change the predicted $r_0$ by more than about a factor of 2. Also,
the directly observed (beam-diluted) brightness temperatures limit the
true $T_{\rm b}$ to be at least $\sim$30\,K considering that the
contribution of image C to A+C is likely minor.

The effective lensing magnification $\mu_{L}^{\rm eff}$ is defined as
the ratio between the observed apparent luminosity and the true
luminosity. It thus is a direct measure for the ratio of the observed
equivalent disk size $r_0$ and the true equivalent disk size $r_0^{\rm
  true}$ via

\begin{equation}
  \mu_{L}^{\rm eff}=(r_0/r_0^{\rm true})^2 .
\end{equation}

This relation is displayed in Fig.~\ref{f17}b for a range of $T_{\rm
  b}$.  The $T_{\rm b}$=50\,K model discussed above is indicated by
the solid line and the boxes (highlighting selected magnification
factors). Using only the given constraints without modeling the system
in more detail, a large range of magnification factors would be in
agreement with the CO data.  However, the impact of changing the
intrinsic $T_{\rm b}$ is relatively minor, as all curves occupy a
narrow range in this plot.

The case of a completely filled, circular CO disk as discussed so far
is a very particular solution, even appreciating the fact that the CO
is likely situated in a relatively compact, rotating circumnuclear
ring. In a more realistic approach, the CO disk has an inner boundary
$r_{\rm in}$ at a certain distance from the central hot nucleus, due
to the simple fact that molecular gas cannot survive at distances
where the ionizing radiation field is strong enough to dissociate the
diatomic CO (and H$_2$) molecules. It thus is a ring rather than a
disk. Also, it is likely that we do not see this ring face-on, but at
a certain inclination $i$ towards the line of sight (LOS). Also, the
molecular material may be distributed in clouds and/or clumps rather
than smoothly, and thus may have an area filling factor $f_{\rm A}$
lower than 1. This can be expressed as (see also Wei\ss\ et
al.~\citeyear{wei07}):

\begin{equation}
r_{\rm true}=\left[\frac{(r_0^{\rm true})^2+r_{\rm in}^2}{f_{\rm A} {\rm cos}\,i
}\right]^{1/2} .
\end{equation}

Note that $i$=0$^\circ$ corresponds to face-on. Also, we use a thin
disk approximation, neglecting the geometrical thickness of the CO
ring. This simplification leads to an overprediction of $r_{\rm true}$
for high inclinations. This approximation is justified, as the high
observed optical brightness of the source in combination with the
detection of large amounts of dust in its circumnuclear torus suggests
that we have a relatively unobscured view on the AGN, and thus that we
likely see the galaxy at a relatively low inclination
($i$$<$35$^\circ$).

In Fig.~\ref{f17}c, a range of inclinations is shown for the $T_{\rm
  b}$=50\,K model in the $\mu_{L}^{\rm eff}$--$r_{\rm true}$ plane.
An inner ring boundary of $r_{\rm in}$=50\,pc is assumed unless stated
otherwise. For comparison, a solution without an inner ring boundary
is shown for $i$=0$^\circ$. The change of the predicted $r_{\rm true}$
is only minor, in particular for the solutions with low
magnifications. In general, the figure shows that the range of
solutions for high inclinations is significantly different from that
for low inclinations. However, in the preferred range of
$0^\circ<i<35^\circ$, the solutions are fairly similar. While $r_{\rm
  in}$ and $i$ can be constrained relatively well by existing
observations, this is not at all true for $f_{\rm A}$. As an example,
a solution with $i$=75$^\circ$ and $f_{\rm A}$=0.5 is shown. The
difference in area filling factor has a large impact on the predicted
solutions. As another example, solutions for $i$=0$^\circ$, $f_{\rm
  A}$=0.5 and $i$=60$^\circ$, $f_{\rm A}$=1.0 are indistinguishable.
The {\em structure} of the molecular ISM thus is an unknown in this
discussion that has significant impact. There are two more interesting
results to note: First, the resolution of our observations and A+C/B
brightness ratio alone set a meaningful lower limit on the
magnification of models with $f_{\rm A} {\rm cos}\,i$ significantly
lower than 1. Second, the models with $f_{\rm A} {\rm cos}\,i$ close
to 1 predict very compact CO rings for high magnifications ($r_{\rm
  true}$=80--125\,pc).

Although the results presented in this subsection significantly
constrain the allowed parameter space for potential lensing models,
they do not allow to strongly constrain the actual magnification.  In
the following, we thus discuss existing lensing models in more detail,
show their limitations, and suggest a new model which overcomes part
of the problems of previous models.

\subsection{Lens Modeling: Previous Models} \label{model_pre}

In most classical lensing models, the surface density of the lensing
galaxy is expressed by an ellipsoidal with a core of finite radius.
However, high-resolution observations of the central regions of
galaxies indicate that the luminosity profiles of some of the most
massive galaxies appear to have central cusps rather than cores with
finite radii (e.g., Faber et al.~\citeyear{fab97}). Such distributions
are reminiscient of the cusp power-law density profiles with a break
radius used for dark matter halos in cosmological simulations (e.g.,
Navarro, Frenk \& White \citeyear{nfw97}, `NFW'), but, in these
observations, are seen for the stellar component of the galaxies.  The
(baryonic) density profile of the lensing galaxy of APM\,08279+5255
has been modeled with both core (Ibata et al.~\citeyear{iba99}; Egami
et al.~\citeyear{e2000}) and cusp (Munoz et al.~\citeyear{mun01};
Lewis et al.~\citeyear{lew02}) configurations previously.

\subsubsection{Cored Singular Isothermal Ellipsoid Models} \label{egami}

The model of Egami et al.\ (\citeyear{e2000}) is a modified singular
isothermal ellipsoid (SIE) model with a finite core. Such models have
six parameters: the Einstein radius, the core radius, the ellipticity
of the lensing potential and its position angle, and the
(2-dimensional) center position of the source. For a three image
source like APM\,08279+5255, observations offer six relevant
constraints for the model: two relative image brightnesses, and two
relative (2-dimensional) image positions. In addition, the fact that
the three images are almost collinear requires the ellipticity to be
close to zero, and thus predicts an almost circular lensing potential.
The geometry of lensing systems requires that any non-singular mass
distribution produces an odd number of lens images (Burke
\citeyear{bur81}). However, if the core is very small, one of the
images is de-magnified, and thus unlikely to be observable (note that
an even number of images is observed toward most lens systems). The
fact that the third image of APM\,08279+5255 is not strongly
de-magnified thus also restricts the core to be non-singular, and thus
to have a rather large radius within this model. Egami et al.\
(\citeyear{e2000}) find a core radius of 0.2$''$, corresponding to
1.2--1.7\,kpc at 0.5$<$$z$$<$3.5, the likely redshift range of the
lensing galaxy, which appears extreme even for the most massive
elliptical galaxies. In particular, quasar image C is located only
0.03$''$ (180--250\,pc) from the center of the lensing potential.  As
the authors already state themselves, one thus might expect
differential reddening of image C in this configuration, which is not
observed. However, the most problematic point of the Egami et al.\
(\citeyear{e2000}) model is that it predicts a rapid change of
morphology with source size (their Fig.~7). For the X-ray to
near-infrared observations, where most of the emission likely comes
from the central $<$0.1--1\,pc region, their model correctly predicts
a three image configuration. This even holds true for a source size of
20\,pc, where the images are elongated, but still clearly seperated,
and thus would be picked up as three pointlike images by observations
at a resolution of, e.g., 0.3$''$. However, for source sizes of 50\,pc
and larger, the images start to form arcs and rings, and ultimately at
220\,pc, a filled sphere. However, in our \aco\ observations, we probe
scales of 100--500\,pc (see above), depending on the lensing
magnification, but still see the 3-image structure as observed in the
X-ray-to-NIR wavelength regime. Moreover, the extended components seen
in radio continuum emission, which likely probes few 100\,pc to kpc
scales, gets lensed into an arc-like structure rather than a filled
sphere (see Fig.~\ref{f1}, right). While this model predicts high
effective lensing magnifications of $\mu_{L}^{\rm eff}$$\sim$100 (and
thus a plausible explanation for the extreme observed properties of
APM\,08279+5255), it is ruled out by the structure detected in our CO
and radio continuum maps.

\begin{figure*}
\includegraphics[angle=-90,scale=0.77]{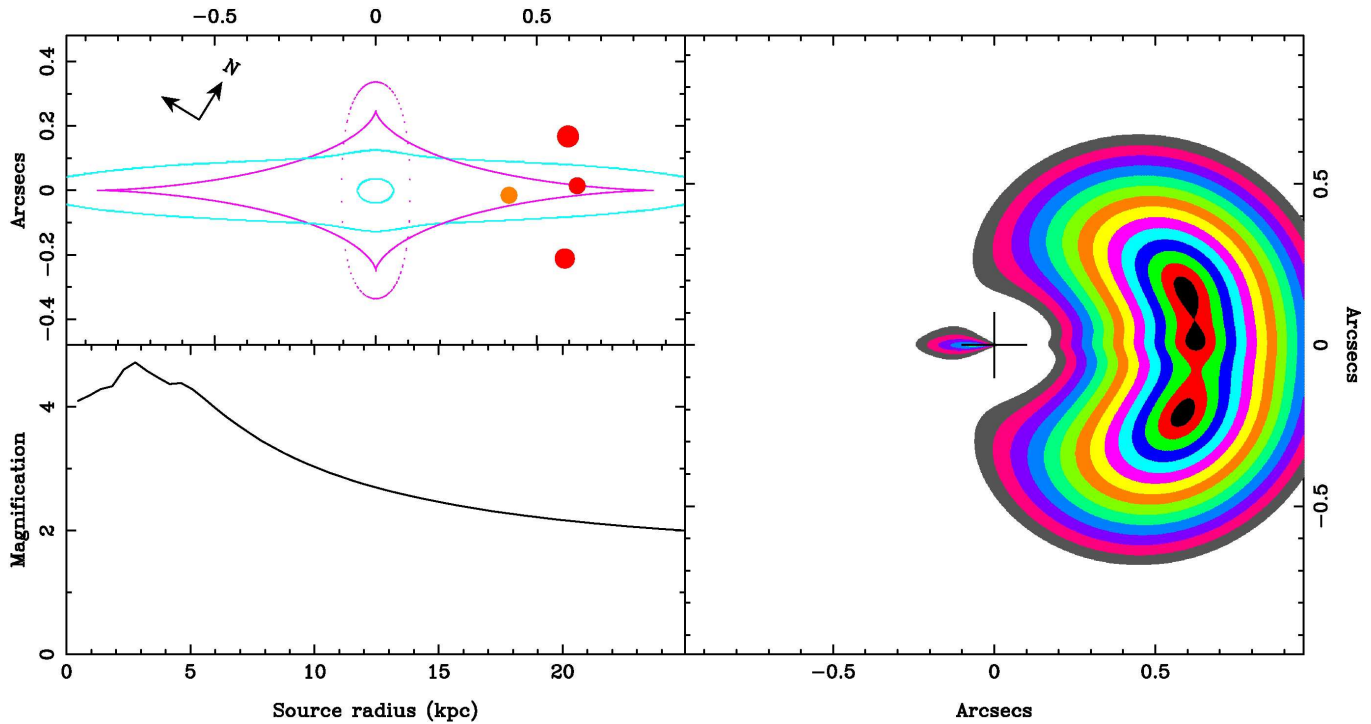}
\caption{ New lensing model for \apm, assuming a highly inclined
  spiral lens.  The top left panel shows the caustic and critical line
  structure. The orange circle indicates the source position, and the
  red circles represent the corresponding image locations. The right
  panel shows the image configuration for a range of source sizes. The
  color contours depict an increase in radius in 0.25\,kpc steps.  The
  bottom left panel shows the corresponding overall magnification as a
  function of radius. In this model, the \aco\ emitting region has a
  true radius of $\sim$0.5\,kpc or less. \label{f16}}
\end{figure*}

\subsubsection{Cusp Models} \label{lewis}

As mentioned above, there is mounting evidence that galaxy cores are
mostly very small, and that their mass distribution possesses a
significant cusp (Rusin \& Ma \citeyear{rm01}; Winn et al.\
\citeyear{win04}). Munoz et al.\ (\citeyear{mun01}) thus suggested a
model complementary to the one by Egami et al.\ (\citeyear{e2000}),
assuming a cusp configuration for the lens in APM\,08279+5255. Instead
of a core radius, such models are parameterized by a cusp power-law
index and a break radius, i.e., 7 parameters in total, and thus are
underconstrained by a three-image lens without further assumptions.
For APM\,08279+5255, cusps with a small break radius are favoured, and
lead to high magnification factors of $\mu_{L}^{\rm eff}$$\sim$100,
similar to the Egami et al.\ (\citeyear{e2000}) model.  However, the
source can only be modeled with relatively shallow cusps with a small
power-law index (0.2--0.4 rather than 1--2 as observed toward most
galaxies). Such shallow cusps are, indeed, relatively similar to cores
with finite radii (corresponding to a power-law index of 0). The
authors conclude that the lens may be a spiral galaxy, since some
spiral galaxies appear to have finite central densities rather than
steep cusps.

Keeton \& Kochanek (\citeyear{kk98}) and Bartelmann \& Loeb
(\citeyear{bl98}) have shown that highly flattened, highly inclined
potentials, as found, e.g., for disks of edge-on spiral galaxies, can
form a `naked cusp' caustic. In such a configuration, the inner
diamond caustic extends outside the elliptical caustic, producing
three roughly collinear images with similar brightness.  The central
third image is not strongly de-magnified, which is different from
typical core and cusp models. However, a generic feature of such
models with a naked cusp in an inclined spiral is that the overall
magnification is significantly lower than in large core lenses.  Such
a peculiar lens configuration, however, is expected to be rare.

Lewis et al.\ (\citeyear{lew02}) have developed a lensing model of the
APM\,08279+5255 system based on such a highly elliptical lens, where
the quasar core of the lensed galaxy is situated in the direct
vincinity of a naked cusp. In their model, the inclined disk of the
lens has a projected axis ratio of 0.25, and a rotational velocity of
200\,\kms. It has a core radius of 0.065\,$h^{-1}$\,kpc, and is
truncated at an outer radius of 8\,$h^{-1}$\,kpc. This truncated,
flattened disk is assumed to sit in a spherical halo.  Due to the
possible detection of microlensing (Lewis et al.\ \citeyear{lew02b};
Chartas et al.\ \citeyear{cha02}), the observed optical and X-ray
brightnesses of the individual images may not reflect the true
relative macrolensing magnifications. Their brightness ratios thus are
not considered a hard constraint for the lens mass model.  Apart from
the optical image positions, Lewis et al.\ (\citeyear{lew02}) also
considered the structure seen in their \aco\ map a constraint for the
lensing model. While the continuum emission underlying the \aco\
emission in their maps is lensed into an arc and aligned with the
optical emission, the \aco\ emission appears significantly more
extended. They thus assumed that the CO disk in APM\,08279+5255 has to
be large enough to reach out beyond the naked cusp and into the
5-image central caustic structure. The resulting lensing magnification
of the optical emission is $\mu_{L}^{\rm opt}$=7, and image brightness
ratios of $b$(A,B)=0.75 and $b$(A,C)=0.75. The model reproduces the
brightness ratio between images A and B (which is not subject to
extreme changes) relatively well, while image C is significantly
brighter than observed. The lensing magnification of the CO emission
in their model is only $\mu_{L}^{\rm CO}$=2.5--3. The true size of the
CO disk thus is 400--1000\,pc, and the CO becomes the dominant
contributor to the total mass within the central kpc of the QSO. Due
to the significantly lower lensing magnification factor compared to
previous studies, this model predicts that APM\,08279+5255 is
intrinsically an extremely bright source at X-ray-to-FIR wavelengths,
rather than being modestly bright.

Our new, improved high-resolution \aco\ maps have revealed that the
extended CO arc found by Lewis et al.\ (\citeyear{lew02}) is a noise
artifact, and that the lensed CO emission is coincident with the
optical images. In contrast to the underlying, slightly more extended
continuum emission, it even shows the image structure seen in the
optical. Assuming that the CO emission is more compact than assumed by
Lewis et al.\ (\citeyear{lew02}) and lies within the region of the
naked cusp, their model describes these new observations quite well.
Their model then suggests that the CO-emitting region has a true
radius of 350\,pc or less, and has a larger magnification factor close
to that of the optical emission. However, this model was derived
taking the extended CO structure found by Lewis et al.\
(\citeyear{lew02}) as a constraint. To find the best-fitting model to
existing observations, it thus is desirable to perform a new parameter
study.

\subsection{Lens Modeling: New Model} \label{model}

In light of our new observations of \apm, we have conducted an
extensive, systematic parameter study of the lensing configurations
allowed within the constraints of its observed properties. All
considered models assume lensing by an inclined spiral galaxy. 

As described above, multiwavelength observations of \apm\ at different
epochs appear to indicate that the image ratios change, possibly due
to microlensing. Thus, they cannot be considered hard constraints for
the primary lens. Leaving the relative image brightnesses as free
parameters, however, would sacrifice two observational constraints,
and, in fact, would under-constrain most lens models. Models
considered good thus are required to reproduce $b$(A,B) relatively
well (due to the relatively minor observed fluctuations), and to
reproduce C as the faintest image.

In Fig.~\ref{f16}, we show the model that gives the overall best fit
to the observed properties at different wavelengths within our study.
The highly inclined disk of the lens in this model has a projected
axis ratio of only 0.09. It is truncated at an outer radius of
15\,$h^{-1}$\,kpc, and has a rotational velocity of 140\,\kms. The top
left panel of Fig.~\ref{f16} shows the caustic and critical line
structure, as well as the position of \apm\ in the source and image
planes. The model predicts $b$(A,B)=0.84, and $b$(A,C)=0.57,
respectively. The brightness ratio of images A and B is close to the
observed value.  The total magnification of a point-like source in
this model is only 4.15, and thus relatively modest. The bottom left
panel shows the differential total magnification as a function of the
source radius. The total magnification scales down by only a factor of
2 from source radii of few pc out to 20\,kpc, showing that
differential lensing effects do not strongly influence the observed
flux ratios.  Out to 5\,kpc, it varies by less than 20\%. The SED
shown in Fig.~\ref{f4a} thus is predicted to not be strongly distorted
by differential lensing effects over the whole spectral range.  The
right panel of Fig.~\ref{f16} shows the observed image configuration
and overall morphology for a range of source sizes. The morphology of
the \aco\ observations is consistent with having a true source radius
of 500\,pc or less in this model. This is consistent with a lensing
factor of 4 in Fig.~\ref{f17}, assuming an inclination of
$<$30$^\circ$ and a high CO area filling factor. In the following, we
will assume $r_{\rm true}$=550\,pc and $\mu_{L}$=4 for the
\aco-emitting region, consistent with both observations and the
lensing model. The model also correctly reproduces the 2.6\,mm
continuum emission underlying the CO line emission, assuming that the
extended continuum emission component comes from a region of about
1\,kpc size.  A more detailed description and analysis of the
uncertainties within this model of the gravitational lensing in \apm\
will be described by G.~Lewis et al.\ (2007, in prep.).

\section{Black Hole, Gas and Dynamical Masses} \label{masses}

\subsection{Black Hole Mass} \label{mbh}

The enormous energy output that powers the huge bolometric luminosity
$L_{\rm bol}$ of APM\,08279+5255 is produced by a central
super-massive black hole (SMBH) that is fed through an accretion disk.
There are several ways to constrain the SMBH mass from observations.
The most basic approach is to assume that the gradient of the
radiation pressure driving gas out of the central region is exactly
counterbalanced by the gravitational force attracting infall of the
same material. This equality of forces provides a limiting mass for
the SMBH. Assuming that dominant mechanism contributing to the opacity
is Thomson scattering off free electrons (Eddington \citeyear{edd21})
leads to:

\begin{equation}
\left(\frac{M_{\rm BH}}{\rm M_\odot} \right)=7.7 \times 10^{-39}
\left(\frac{L_{\rm Edd}}{\rm erg\,s^{-1}} \right) .
\end{equation}

Assuming $L_{\rm bol}$=$L_{\rm Edd}$ gives $M_{\rm BH}^{\rm Edd}$=2.1
$\times$ 10$^{11}\,\mu_{L}^{-1}$\,M$_\odot$.

Another, perhaps better constrained estimate is provided by
observations of the 2$s^2S_{1/2}$--2$p^2P_{1/2,3/2}$ resonance
transitions of the C$^{3+}$ ion (\civ$\lambda\lambda$1548,1550).  From
reverberation mapping of nearby active galactic nuclei, Vestergaard \&
Peterson (\citeyear{vp06}) have found an empirical relation between
the \civ$\lambda\lambda$1548,1550 FWHM linewidth $\Delta v_{\rm
  FWHM}^{\rm C\ts {\scriptsize IV}}$, the rest-frame UV continuum
luminosity at 1350\,\AA, and the mass of the central black hole:

\begin{equation}
\left(\frac{M_{\rm BH}}{\rm M_\odot} \right)=4.6 \times
\left(\frac{\Delta v_{\rm FWHM}^{\rm C\ts {\scriptsize IV}}}{\rm
km\,s^{-1} } \right)^2\,\left(\frac{\lambda L_\lambda (1350\,\rm \AA
)}{10^{44}\,\rm erg\,s^{-1}} \right)^{0.53} .
\end{equation}

From the spectrum by Irwin et al.\ (\citeyear{irw98}), we thus derive
$M_{\rm BH}^{\rm C\ts {\scriptsize IV}}$=9.0 $\times$
10$^{10}\,\mu_{L}^{-1}$\,M$_\odot$, or about half of the Eddington
limit. However, note that the reverberation-based masses underlying
this equation are uncertain by a factor of about 3 (Vestergaard \&
Peterson \citeyear{vp06}). Also, part of the wings of the \civ\ line
in APM\,08279+5255 are subject to strong absorption, making the
derivation of the linewidth somewhat uncertain. However, the estimated
width of the \civ\ line is similar to that of the Pa$\alpha$ and
Pa$\beta$ lines of hydrogen (Soifer et al.~\citeyear{soi04}).
Finally, the accretion disk may be inclined toward the line of sight,
which would also influence this result.

\subsection{Molecular Gas Mass} \label{mgas}

To derive molecular gas masses $M_{\rm gas}$ from CO luminosities in
high redshift quasars, it currently is the most common approach to
adopt a well-established conversion factor $\alpha_{\rm CO}$ for
nearby ULIRGs ($\alpha_{\rm
  CO}$=0.8\,M$_\odot$(K\,\kms\,pc$^2$)$^{-1}$; Downes \& Solomon
\citeyear{ds98}).  This is justified by the fact that the observed
brightness temperatures in distant quasars appear to be similar to
those of local ULIRGs. However, $\alpha_{\rm CO}$ also depends on the
gas density, which appears to be much higher in APM\,08279+5255.
Wei\ss\ et al.\ (\citeyear{wei07}) thus re-derive a much higher
conversion factor of $\alpha_{\rm CO}$ $\simeq$
5\,M$_\odot$(K\,\kms\,pc$^2$)$^{-1}$ for this source.  From our \aco\
luminosity, we thus derive $M_{\rm gas}$=5.3 $\times$
10$^{11}\,\mu_{L}^{-1}$\,M$_\odot$. This corresponds to about six
times $M_{\rm BH}$ (based on the \civ\ estimate), assuming that no
significant differential lensing effects act between the scales of the
SMBH and of the molecular gas reservoir. This is somewhat lower than
the factor of roughly ten found by Shields et al.\ (\citeyear{shi06})
for a larger sample of distant gas-rich quasars, but in agreement
within the large spread of observed values.

\subsection{Dynamical Mass} \label{mdyn}

From Newtonian mechanics, a dynamical mass can be derived for the
rotating molecular disk in APM\,08279+5255 (e.g., Solomon \& Vanden
Bout \citeyear{sv05}):

\begin{equation}
\left(\frac{M_{\rm dyn}}{\rm M_\odot} \right)=2.34 \times 10^5
\left(\frac{r_{\rm true}}{\rm kpc} \right)\,\left(\frac{\Delta v_{\rm
FWHM}^{\rm CO}}{\rm km\,s^{-1} } \right)^2\,{\rm sin}^{-2} i ,
\end{equation}

where $\Delta v_{\rm FWHM}^{\rm CO}$ is the observed CO linewidth, and
$r_{\rm true}$ is the derived true radius of the CO disk. Using the
observed \aco\ linewidth of 556 $\pm$ 55\,\kms\ (Riechers et
al.~\citeyear{rie06}), and using $r_{\rm true}$=550\,pc, we derive
$M_{\rm dyn}$\,sin$^2 i$ = 4.0 $\times$ 10$^{10}$\,M$_\odot$.  This
dynamical mass of the central region of APM\,08279+5255 corresponds to
the sum of black hole mass, gas mass, dust mass and stellar mass (plus
dark matter).  Assuming a gas-to-dust ratio of 150 (Wei\ss\ et
al.~\citeyear{wei07}), the contribution of dust to the total mass
budget is negligible. For the moment, we will also assume that the
mass fraction of the stellar bulge in the central 550\,pc is small,
i.e., that $M_{\rm dyn}$\,sin$^2 i$$\simeq$($M_{\rm gas}$+$M_{\rm
BH}$).  Assuming a face-on, unlensed disk, $M_{\rm gas}$+$M_{\rm BH}$
would exceed the dynamical mass by a factor of 15.5, and thus is
ruled out by the observations.  Assuming $\mu_{L}$=4, we thus find
that the CO disk in APM\,08279+5255 can be inclined by at most
$i$=30$^\circ$ to not exceed the dynamical mass. With $\Delta v_{\rm
rot}$=$\Delta v_{\rm FWHM}^{\rm CO}$\,sin$^{-1} i$, this implies a
rotational velocity of 1095\,\kms.

\subsection{Stellar Bulge Mass Limit} \label{mbulge}

As mentioned in the previous section, there is a third important
constituent to the dynamical mass, i.e., the mass of the stellar
bulge. This is a quantity that currently cannot be observed in the
highest redshift galaxies (particularly in quasars, where the AGN
outshines any stellar light from the host). However, from our
observations and theoretical considerations, we can derive a limit for
the mass fraction of the stellar bulge within $r_{\rm true}$. Carilli
\& Wang (\citeyear{cw06}) find that high-redshift gas-rich QSOs have a
median $\Delta v_{\rm FWHM}^{\rm CO}$ of 300\,\kms, at a mean
inclination of $\langle i \rangle$=13$^\circ$ (see their erratum).
This corresponds to a mean rotational velocity of $\langle\Delta
v_{\rm rot}\rangle$=1335\,\kms. Assuming that this value applies to
APM\,08279+5255, we derive an inclination of $i$=25$^\circ$. Assuming
$M_{\rm dyn}$\,sin$^2 i$=($M_{\rm gas}$+$M_{\rm dust}$+$M_{\rm
BH}$+$M_{\rm bulge}$), and a gas-to-dust ratio of 150, this
corresponds to a bulge mass of $M_{\rm bulge}$=3.0 $\times$
10$^{11}\,\mu_{L}^{-1}$\,M$_\odot$, or $M_{\rm bulge}$=3.4\,$M_{\rm
BH}$ (0.57\,$M_{\rm gas}$; see also Tab.~\ref{t5}) for $\mu_{L}$=4.
For comparison, if a model with $\mu_{L}$=100 (and thus $r_{\rm
true}$=125\,pc) that fits the data were to exist, it would give
$M_{\rm bulge}$=4.4 $\times$ 10$^{12}\,\mu_{L}^{-1}$\,M$_\odot$, or
$M_{\rm bulge}$=50\,$M_{\rm BH}$ (8.4\,$M_{\rm gas}$). This means that
for low-$\mu_{L}$ models, the gas component dominates the total
baryonic mass budget in the central region of \apm, while for
high-$\mu_{L}$ models, the stellar bulge is the dominant contributor.

\begin{deluxetable}{lccc}
\tabletypesize{\scriptsize}
\tablecaption{Mass Budget in the Central Region of APM\,08279+5255.\label{t5}}
\tablehead{
Component & Mass & & Mass Fraction\\
 & lensed & unlensed & \\
 & [10$^{10}$\,($\mu_{L}^{-1}$)\,M$_\odot$] & [10$^{10}$\,M$_\odot$] & [\%] }
\startdata
Molecular Gas & 53 & 13 & 57.4 \\
Dust & 0.35 & 0.09 & 0.4 \\
Black Hole & 9.0 & 2.3 & 9.7 \\
Stellar Bulge\tablenotemark{a} & 30 & 7.5 & 32.5 \\
\vspace{-3mm}
\enddata
\tablenotetext{a}{Fraction within the radius of the CO disk.}
\tablecomments{Here, we assume $i$=25$^\circ$, $r_{\rm
true}$=550\,pc, and $\mu_{L}$=4. Masses are uncorrected/corrected for lensing,
assuming $\mu_{L}$=4.}
\end{deluxetable}

\subsection{Limitations of this Analysis} \label{limitations}

The maximum CO velocity widths predicted by the Carilli \& Wang
(\citeyear{cw06}) unified model exceed the rotational velocities of
the most rapidly rotating nearby giant ellipticals by about a factor
of 2. This may suggest that the observed linewidths in the systems
with broader but not multiply peaked CO lines are not due to
rotationally supported disks, but possibly in a later stage of a major
merger where the disk is not self-gravitating yet. In that case, the
mean inclination of quasar host galaxies derived by Carilli \& Wang
may overpredict the rotational velocities for gravitationally bound
molecular disks.  This is of particular concern for \apm, as its CO
velocity width is already the highest observed among the distant
quasars. If the measured CO linewidth was that of a gravitationally
bound disk and representative for the rotational velocity inside a
present day giant elliptical, the inclination could not be much lower
than $i$=60$^\circ$.  This leaves a factor of 2--3 discrepancy in the
$M_{\rm dyn}$ estimate relative to the $i$=30$^\circ$ solution.

The uncertainties in the luminosity-based mass estimates are a factor
of a few. Together with the uncertainties in disk size and lensing
factor, and the contribution of dark matter to the mass budget, the
real inclination of the molecular disk remains uncertain (as only a
factor of 3--4 difference exists between high and low inclination
solutions). Due to the degeneracy of area filling factor $f_A$ and
inclination, our results thus are still in agreement with those of
Wei\ss\ \etal\ (\citeyear{wei07}) within conservative error estimates,
even with the new morphological constraints.

\section{On the Radio--Far-Infrared Correlation} \label{radio-FIR}

In the local universe, star-forming galaxies follow a tight
correlation between the monochromatic radio continuum luminosity at
1.4\,GHz and the far-infrared luminosity ($L_{\rm FIR}$), which holds
over more than 4 orders of magnitude (e.g., Condon \etal\
\citeyear{con91}; Condon \citeyear{con92}; Yun \etal\
\citeyear{yun01}; Bell \citeyear{bell03}). This correlation is
probably due to coupling between infrared thermal dust emission due to
heating by young stars and non-thermal radio synchrotron emission
from supernova explosions and SNRs (e.g., Helou \etal\
\citeyear{hel85}), and constitutes the basis for the Condon model
described above. The radio--FIR correlation can be expressed via the
so-called $q$-parameter (Helou \etal\ \citeyear{hel85}):

\begin{equation}
q={\rm log}\left(\frac{L_{\rm FIR}}{9.8 \times 10^{-15}\,{\rm L_\odot}}\right)-{\rm log}\left(\frac{L_{\rm 1.4\,GHz}}{{\rm W\,Hz^{-1}}}\right) .
\end{equation}

The apparent FIR luminosity of \apm\ is $L_{\rm FIR}$=(2.0 $\pm$ 0.5)
$\times$ 10$^{14}\,\mu_{L}^{-1}$L$_\odot$ (Beelen \etal\
\citeyear{bee06}; Wei\ss\ \etal\ \citeyear{wei07}). We thus obtain
$q$=2.03, a value which falls well within the range of values found
for IRAS galaxies (Yun \etal\ \citeyear{yun01}), and consistent with
the values found for other dust and gas-rich high-$z$ quasars (Carilli
\etal\ \citeyear{car04}; Beelen \etal\ \citeyear{bee06}; Wang \etal\
\citeyear{wan07}). A common interpretation of a $q$ value close to the
radio-FIR correlation is that star formation is the dominant process
to power both the radio and FIR emission. However, using the dust
model of Wei\ss\ \etal\ (\citeyear{wei07}), our analysis of \apm\ has
shown that the AGN may power the dominant fraction of both the FIR and
the radio emission. We obtain this conclusion based on the fact that
the SED of \apm\ cannot be fitted with a simple Condon model (which
motivates the star formation interpretation of the radio-FIR
correlation). We thus conclude that caution has to be used when
interpreting the properties of dust and gas-rich high-$z$ quasars
based on the radio-FIR correlation alone.

\section{Discussion}

The observed properties of \apm\ are undoubtedly peculiar. It thus is
important to understand whether this peculiarity is due to a rare lens
configuration, possibly leading to very high lensing magnification
factors and significant differential lensing effects, or whether \apm\
is a source that is just moderately magnified by gravitational lensing
and thus intrinsically extreme.

\subsection{Multiwavelength Properties of \apm}

\apm\ shows broad absorption lines in the optical/UV, and emission
from high excitation lines like \civ\ and \nv\ close to the central
nucleus (Irwin \etal\ \citeyear{irw98}), kinematically blueshifted
relative to the molecular gas and \ci\ emission by about 2500\,\kms\
(Downes \etal\ \citeyear {dow99}; Wagg \etal\ \citeyear{wag06}). The
quasar also shows relativistic X-ray broad absorption lines from even
higher emission lines of iron from gas situated within the UV BAL
region (Chartas \etal\ \citeyear{cha02}; Hasinger \etal\
\citeyear{has02}).  The BALs come from an outflow of highly ionized
gas from the accretion disk, mostly driven by radiation pressure. This
outflow distributes the accretion disk material over the central
region of the quasar, and out into the host galaxy. The X-ray BAL
region is likely a dust-free, high column density absorber responsible
for shielding, as indicated by the detection of an iron K-shell
absorption edge (Hasinger \etal\ \citeyear{has02}). This shielding is
responsible for the typical X-ray faintness of BAL QSOs.  The X-ray
luminosity alone thus is not a good measure for the energy output of
\apm. The detection of strong iron lines also indicates that the X-ray
BAL emission comes from significantly metal-enriched gas with possibly
super-solar metallicities. While this strongly indicates that the gas
has already been processed in a starburst cycle, it is biased toward
the very central region (possibly $<$0.1\,pc). It thus cannot be
assumed ad hoc that the metallicity is super-solar over the whole
scale of the galaxy. The continuum X-ray emission in \apm\ has a
photon index that is in qualitative agreement with a radio-quiet
quasar (RQQ). The optical/IR shape of the SED is also consistent with
\apm\ being a RQQ. Even more so, it follows the radio-quiet locus of
the radio--FIR correlation, which indicates that the copious amounts
of dust in this source may be heated to a significant fraction by
newly formed young stars, while the radio synchrotron emission may
originate from supernova remnants (Irwin \etal\ \citeyear{irw98};
Beelen \etal\ \citeyear{bee06}). However, our analysis of the radio
and dust properties indicates that most of the radio/FIR emission may
well be AGN-related.  Within the unified theory of AGN galaxies, the
smoothness of the IR spectrum, in particular the lack of a pronounced
IR peak, indicates that the accretion disk in this radio-quiet quasar
is seen close to face-on (Soifer \etal\ \citeyear{soi04}). This agrees
well with the enormous optical brightness and the flatness of the
optical spectrum (Irwin \etal\ \citeyear{irw98}; Egami \etal\
\citeyear{e2000}), indicating a lack of obscuration and thus an angle
of view which clearly lies within the ionization cone.  The hot dust
in the parsec-scale accretion disk is heated by the central AGN, and
extends outwards into a warm dusty torus, extending out to 100\,pc
scales.

The luminous AGN in \apm\ is the dominant power source out to large
scales, which even may generate a large fraction of the huge
(far-)infrared luminosity in this source by heating the
dust.\footnote{It is important to note that, due to the unique
properties of \apm, additional diagnostics are desirable to
undoubtedly disentangle the heating sources of the gas and dust.}  The
apparent FIR luminosity of \apm\ is $L_{\rm FIR}$=(2.0 $\pm$ 0.5)
$\times$ 10$^{14}\,\mu_{L}^{-1}$L$_\odot$ (Beelen \etal\
\citeyear{bee06}; Wei\ss\ \etal\ \citeyear{wei07}).  Wei\ss\ \etal\
(\citeyear{wei07}) have found that about 90\% of the FIR luminosity
are due to emission from warm 220\,K dust which may be powered by the
AGN, while only 10\% are due to colder 65\,K dust that is likely
heated by star formation.  This would explain why a significant
fraction of the dust in \apm\ on extended (100s pc) scales is warmer
than in typical ULIRGs, or even other high-$z$, dust-rich QSOs. It
would also explain why \apm\ is an outlier on the $L'_{\rm
CO}$--$L_{\rm FIR}$ relation (Riechers
\etal\ \citeyear{rie06}), which is set by galaxies where the FIR
luminosity is dominated by heating from star formation.  Subtracting
the 90\% of $L_{\rm FIR}$ that may be powered by the AGN, \apm\
follows the $L'_{\rm CO}$--$L_{\rm FIR}$ relation quite well.
However, note that even though the apparent FIR luminosity of \apm\ is
extreme, this would indicate that only 3\% of the total quasar power
(as measured by $L_{\rm bol}$) are re-emitted in the FIR wavelength
regime (assuming no differential lensing effects play into the ratio).
We find that up to 80--90\% of the radio luminosity in \apm\ may be
powered by the AGN, and only 10--20\% by the starburst. Due to the
fact that this ratio is similar to that found for the FIR luminosity,
\apm\ still follows the radio--FIR correlation for star-forming
galaxies.

\apm\ thus is a radio-quiet, optically luminous dust-rich quasar in
which the central accretion disk and dust torus are likely seen close
to face-on (note however the above concern about the inclination). Its
high FIR luminosity can be explained by dust heating of the AGN, as
can be the high temperature of the dust. It even explains the high CO
excitation if the molecular gas is situated in a rather compact (and
thus dense) circumnuclear ring seen at a low inclination toward the
line of sight, just like the central region. The high dust and gas
temperatures and relatively high densities also explain the
exceptional HCN (and HCO$^+$) excitation in this system, as IR-pumping
via the $\nu_2$=1 vibrational bending mode becomes efficient at such
high temperatures (Wei\ss\ \etal\ \citeyear{wei07}).  Without
shielding and self-shielding, the gas and dust would be even warmer
than observed. Assuming that only the fraction of $L_{\rm bol}$
re-radiated in the FIR reaches out to the molecular regions, the
temperature at a radius of 125\,pc would still be 200\,K. Even at
550\,pc, the temperature would not drop significantly below 100\,K.
The LVG-predicted temperatures for the cool, dense gas component
($T_{\rm kin}=65$\,K) thus require self-screening if the gas is indeed
located in a 550\,pc region around the AGN.\footnote{Due to the fact
  that $T \propto \mu_L^{-1/4}$ and $\mu_L \propto r_{\rm true}^2$,
  the $\mu_L$=4 model requires more screening than the previous
  $\mu_L$=100 models.} Note that all the above may be considered
extreme (i.e., rarely observed). However, all of these effects can be
explained physically without assuming extreme and/or differential
lensing effects. The CO luminosity in \apm\ is relatively modest
(compared to other high-$z$ sources) even without correcting for
lensing. If $L_{\rm FIR}$ is corrected assuming a lensing
magnification factor of only 4, as suggested by our new model, it is
not the highest observed value even without correcting for AGN
heating. Both $L'_{\rm CO}$ and $L_{\rm FIR}$ are higher for the
$z$=4.7 quasar BR\,1202-0725, which is thought to be unlensed, but
undergoing a massive, gas-rich merger event\footnote{Both the CO and
  FIR emission in BR\,1202-0725 are however distributed over at least
  two distinct components, which then each have comparable but lower
  luminosities than \apm.} (Carilli \etal\ \citeyear{car02}; Riechers
\etal\ \citeyear{rie06}).

\subsection{The Black Hole in \apm}

Correcting for $\mu_{L}$=4 predicts $M_{\rm BH}$=2.3 $\times$
10$^{10}$\,M$_\odot$ for \apm. This appears to be extreme compared to
more typical SMBH masses of a few times 10$^9$\,M$_\odot$ of high-$z$
quasars.  However, comparable or even higher black hole masses are
found for BRI\,1335-0417 ($z$=4.41) and BR\,1202-0725 (however, the
estimates for these sources are based on strongly absorbed \civ\
emission lines; Storrie-Lombardi \etal\ \citeyear{sl96}; Shields
\etal\ \citeyear{shi06}). Moreover, Vestergaard (\citeyear{ves04}) has
found that for the range of black hole masses, an upper envelope of
$M_{\rm BH}$$\sim$10$^{10}$\,M$_\odot$ is observed over a large range
in redshift (4$\lesssim$$z$$\lesssim$6 for her sample). So, again,
while \apm\ appears relatively extreme, even its AGN properties are
far from being unique in a scenario assuming only $\mu_{L}$=4. The
extreme optical luminosity may then be explained by a luminous AGN
episode of such a few times 10$^{10}$\,M$_\odot$ SMBH, which accretes
at (super-)Eddington rates (within the uncertainties of the BH mass
estimators).  Also, the formation of such a SMBH at the high mass end
of the BH mass function by a redshift of 4 is consistent with
`downsizing', which predicts that such high mass black holes form at
the high density peaks at high redshift, and are built up rapidly by
vigorous accretion (e.g., Di Matteo \etal\ \citeyear{mat07}). Due to
the fact that the SMBH in \apm\ is among the most massive black holes
that are observed, it is likely to end up as a central cD galaxy of a
massive cluster.  For such galaxies, recent cosmological simulations
suggest that a large fraction of the mass of the black hole grows from
mergers with other black holes rather than being accreted by a single
progenitor black hole (Sijacki \etal\ \citeyear{sij07}).

\subsection{The Stellar Bulge of \apm}

In addition to $M_{\rm BH}$, we were also able to derive a limit on
the mass of the stellar bulge ($M_{\rm bulge}$) for \apm. However, our
dynamical mass estimate, which is used as the predictor for $M_{\rm
  bulge}$, is only valid for the inner region of the galaxy where the
bulk of the molecular gas is found. For the $\mu_{L}$=4 model, this
describes a region with a radius of 550\,pc. For galaxies with $M_{\rm
  BH}$$>$10$^9$\,M$_\odot$ used in local studies of the $M_{\rm
  BH}$--$\sigma_{\rm bulge}$ relation, it has been found that the host
galaxies are typically giant elliptical galaxies with effective radii
of 1.5--8\,kpc (Tremaine \etal\ \citeyear{tre02}; Faber \etal\
\citeyear{fab97}). Assuming that the volume density of stellar
luminosity (and thus mass) flattens with radius from $r^{-2}$ to
$r^{-1}$ in the inner region of the elliptical host galaxy (Gebhardt
\etal\ \citeyear{geb96}), or even that it flattens into a `core', a
significant fraction of the bulge mass is expected to be found inside
the inner 550\,pc. From the local $M_{\rm BH}$--$M_{\rm bulge}$
relationship, it is expected that $M_{\rm bulge}$$\sim$700\,$M_{\rm
BH}$. For the inner region of \apm, we have found $M_{\rm
bulge}$=3.4\,$M_{\rm BH}$. Although uncertain within a factor of a
few, this value falls short by more than two orders of magnitude
compared to the local estimate. Even if corrected for a 10\,kpc radius
host galaxy, a clear offset from the local $M_{\rm BH}$--$M_{\rm
bulge}$ relationship remains\footnote{Note that the local $M_{\rm
BH}$--$\sigma_{\rm bulge}$ relationship is difficult to reconcile even
with a $\mu_{L}$=100 model, which leaves more room for $M_{\rm bulge}$
within $M_{\rm dyn}$ (i.e., $M_{\rm bulge}$=50\,$M_{\rm BH}$), but
also has to assume host galaxy size relations for less massive black
holes.}. This is consistent with what is found for other quasars at
even higher redshift studied with the same technique (Walter \etal\
\citeyear{wal04}; Riechers et al.\ \citeyear{rie08a}, \citeyear{rie08b}; 
see also discussions by Shields \etal\ \citeyear{shi06}; Ho
\citeyear{ho07}). Based on different techniques, similar trends are
found for various galaxy samples at low and intermediate redshift
(e.g., Peng et al.\ \citeyear{pen06}; McLure et al.\ \citeyear{mcl06};
Woo \etal\ \citeyear{woo06}; Salviander et al.\ \citeyear{sal07}).
This may indicate a breakdown of the local $M_{\rm BH}$--$M_{\rm
bulge}$ relationship toward high redshift, and that the mass
correlation between SMBHs and their stellar bulges is not a universal
property, but rather an endpoint of a long-lasting evolutionary
process throughout cosmic times.  We thus conclude that the SMBH in
\apm\ appears to already be largely in place, while the buildup of the
stellar bulge is still in progress.

\subsection{A Possible Evolutionary Scenario for \apm}

If the assumptions made to derive the properties of \apm\ are correct,
this result has several interesting implications. It appears as if
SMBH growth and formation through gas dissipation and accretion (and
possibly mergers) was the main characteristic during the early
formation of this source (even relative to star formation).  Since the
SMBH in \apm\ is already found at the high mass end of the black hole
mass function, it is expected that most of the molecular gas present
at $z$=3.9 is either blown out or formed into stars by $z$=0, while
only a minor fraction is accreted onto the black hole.  Moreover, if
\apm\ is to evolve into a galaxy that fulfills the $M_{\rm 
 BH}$--$\sigma_{\rm bulge}$ relation by $z$=0, a significant fraction
of the molecular gas has to actually end up in stars. In fact, even if
the whole amount of molecular gas that is currently present in the
host galaxy of \apm\ were to be converted into stars by $z$=0, it
would not be sufficient to reach the local $M_{\rm BH}$--$\sigma_{\rm
bulge}$ relation. If our picture of this galaxy is correct, this
result implies that the buildup of the stellar bulge cannot be
accomplished by only converting the observed molecular gas reservoir
into stars, but relies to a significant part on the accumulation of
stellar matter via other mechanisms, and/or the the accretion of
additional star-forming material. Although further details are
difficult to quantify at present, it thus is likely that a significant
fraction of the spheroid will be produced via major and minor mergers
(which are also required for the re-distribution of angular momentum).
It has been suggested that a common, growth-limiting feedback
mechanism acting on both star formation and black hole assembly is the
physical explanation for the local $M_{\rm BH}$--$\sigma_{\rm bulge}$
relation.  Assuming that such a mechanism exists, quasar winds as a
form of AGN feedback would be an obvious candidate for a system like
\apm\ (Silk \& Rees \citeyear{sr98}). One difficulty with this
assumption would then be that the SMBH in this source already accretes
radiatively highly efficiently at (super-)Eddington rates (see above),
where such feedback effects to regulate the further growth of the SMBH
are expected to be very strong. If this effect were to also regulate
star formation in the host galaxy, it would be expected to shut off
both on a relatively short timescale by pushing the remaining gas
outwards.  The build-up of the stellar bulge then can only proceed
further through hierarchical merging and/or accretion of additional
gas, which is expected to again feed both the SMBH and star formation.
To end up anywhere near the local $M_{\rm BH}$--$\sigma_{\rm bulge}$
relation through a self-regulated mechanism, \apm\ thus is required to
form and/or accumulate stars more efficiently (relative to the black
hole feeding/merger rate) than in the past. One possibility to build
up stellar mass while avoiding significant SMBH growth may be through
subsequent `{\em dry}' mergers (e.g., Rines et al.\ \citeyear{rin07}),
which may act in combination with other processes to assemble the
massive host of \apm\ by $z$=0.

\subsection{Our Current Picture of \apm}

There are two main points that support the $\mu_{L}$=4 lensing model
proposed for \apm.  First, the observed \aco\ properties give rather
tight constraints on the allowed combinations of intrinsic CO disk
size and magnification strength for the favoured range of low disk
inclinations, assuming a relatively high CO area filling factor. The
$\mu_{L}$=4 model lies well within the range of these constraints.
Second, the spatially resolved multiwavelength observations of \apm\
show similar morphologies of the lensed source from the X-ray to radio
wavelength regime, probing sub-pc to hundreds of pc scales, while
showing a different morphology out to kpc scales. The $\mu_{L}$=4
model produces this quite naturally, predicting modest effects of
differential lensing. The high lensing magnification models suggested
so far do not reproduce the second characteristic. Also, for low
inclinations, they predict a very compact CO disk of only about
100\,pc radius. This is by a factor of a few less than in typical
ULIRGs (Downes \& Solomon \citeyear{ds98}). The $\mu_{L}$=4 model, on
the other hand, produces a CO disk of 550\,pc radius, which is in a
more typical regime.

While some properties of \apm, like the optical luminosity and
expected strong tidal forces from the black hole, as well as the
requirement of significant shielding of the molecular gas, may still
favour a scenario in which the source is highly magnified by
gravitational lensing, there thus are several arguments that favour
models with low $\mu_{L}$. It however is important to note that, if a
high-$\mu_{L}$ model was to be found that fits the observed
morphological properties of \apm\ (see Krips et al.\ \citeyear{kri07}
for a recent attempt), further studies would be necessary to exclude
one scenario or the other.  Based on the existing observations, no
definitive conclusion can be drawn -- the final decision can only be
made by detecting the lens itself. The lensing model presented in this
paper aids in developing a quite consistent picture of \apm\ based on
the existing observations, describing the source as a dust and
gas-rich galaxy with a very massive, active black hole. The deep
gravitational potential of the central region of this source causes
the molecular gas to be rather dense, while the strong, penetrating
AGN radiation causes it (and the dust) to be rather warm.  The whole
system may be seen relatively close to face-on (the inclination is
however difficult to constrain).  It has a co-eval starburst which
only contributes about 10\% to the FIR dust heating, but still
produces more than 500\,M$_\odot$\,yr$^{-1}$ (Wei\ss\ \etal\
\citeyear{wei07}, correcting for $\mu_{L}$=4). The star formation in
this galaxy takes place far off the black hole mass--bulge mass
scaling relation of nearby galaxies, suggesting that the latter may be
subject to significant evolution throughout cosmic times.

\section{Summary}

In this paper, we present improved, high resolution \aco\ observations
toward the $z$=3.9 BAL quasar \apm\ with the VLA. We also present
multiwavelength radio continuum observations, and a search for \acn\
emission. The source's properties are investigated based on a revised
model of the gravitational lensing in this system. 

From our comprehensive analysis, we obtain the following main results:

1. The \aco\ emission is resolved and has a (lensed) source size of
$\sim$0.3$''$. The structure shows a morphology that is very similar
to the X-ray/optical/NIR morphology, indicative of three images of a
compact, gravitationally lensed source (of which two are blended in
the CO maps). In particular, we do not find any evidence for extended
\aco\ emission over a scale of several arcseconds, in contrast to
previous reports (cf.\ Papadopoulos \etal\ \citeyear{pap01}). The
whole single-dish \aco\ flux found by Riechers et al.\
(\citeyear{rie06}) is recovered from the compact structure detected in
our VLA maps.  The rest-frame 2.6\,mm continuum emission underlying
the \aco\ line is slightly more extended than the CO source, and may
correspond to differentially lensed emission either due to free-free
radiation from the starburst or extended AGN-related emission (in
agreement with the overall steep spectral index). A radio source is
detected 3$''$ south of \apm\ in two out of six radio continuum bands
(see also Ibata \etal\ \citeyear{iba99}), but probably not related to
the high redshift quasar.

2. We derive an apparent \aco\ luminosity of $L'_{\rm CO(1-0)} = (10.6
\pm 0.9) \times 10^{10}\,\mu_{L}^{-1}$ K\,\kms\,pc$^2$.  This is by
about 20\% lower than the re-derived \bco\ luminosity, and even by
40\% lower than the \dco\ luminosity. This may indicate that the
optical depth in the lower $J$ CO transitions is only modest, as
already indicated by previous studies (Riechers \etal\
\citeyear{rie06}; Wei\ss\ \etal\ \citeyear{wei07}). Due to the high
gas density, the \aco\ luminosity corresponds to a total apparent
H$_2$ molecular gas mass of $M_{\rm gas}$=5.3 $\times$
10$^{11}\,\mu_{L}^{-1}$\,M$_\odot$.

3. We set a limit on the \acn\ luminosity in this source (based on GBT
spectroscopy), which is $<$37\% of the \aco\ luminosity.

4. The radio continuum SED of \apm\ is likely dominated by AGN
emission. Based on a starburst galaxy model (Condon \citeyear{con92}),
we conclude that free-free and synchrotron emission from the host
galaxy contribute $\lesssim$20\% to the radio continuum emission (note
that this fraction is clearly model-dependent).  We also find that
the AGN lies in the transition region between radio-quiet and
radio-loud quasars, and is in overall agreement with being
radio-quiet. The low frequency part of the radio SED is
(ultra-)steep, which is commonly found in high-$z$ quasars with
extended radio emission. The overall spectral slope of the radio SED
is in agreement with synchrotron emission that may originate from both
the starburst and the AGN.

5. In spite of the fact that both the FIR (Wei\ss\ \etal\
\citeyear{wei07}) and radio continuum emission in \apm\ appear to be
dominantly powered by the AGN, it follows the radio--FIR correlation
for star-forming galaxies without correcting for the AGN contribution
to the radio/FIR luminosities ($\sim$80--90\% in both cases). This may
indicate that loyalty to the radio--FIR correlation does not
necessarily imply that a galaxy's emission is star formation dominated
on an object-to-object basis.

6. From our CO maps, we derive observed brightness temperatures of
$T_{\rm b}^{\rm obs,A+C}$$\simeq$23\,K for the blend of optical quasar
images A+C and $T_{\rm b}^{\rm obs,B}$$\simeq$13\,K for quasar image
B.  Assuming a true brightness temperature of 50\,K as predicted by
LVG models of the CO excitation (Wei\ss\ \etal\ \citeyear{wei07}),
this corresponds to a lensed CO equivalent disk with a radius of
$\sim$900\,pc.

7. Our revised lensing model of \apm\ predicts that the luminosity is
only enhanced by a lensing magnification factor of about $\mu_{L}$=4
over the whole observed wavelength range. Assuming that the CO disk is
seen relatively close to face-on ($i$$<$30$^\circ$) and has a large
area filling factor, this model is consistent with a CO ring that has
an unlensed radius of about 550\,pc.

8. From \civ\ observations of \apm\ and a scaling relation of nearby
galaxies from the literature, we derive an apparent black hole mass of
$M_{\rm BH}$=9.0 $\times$ 10$^{10}\,\mu_{L}^{-1}$\,M$_\odot$, which
corresponds to about half of the Eddington limit.  This also
corresponds to about 17\% of the molecular gas mass.

9. From our CO maps and model, we derive a dynamical mass of $M_{\rm
  dyn}$\,sin$^2 i$ = 4.0 $\times$ 10$^{10}$\,M$_\odot$ for \apm.
Assuming $\mu_{L}$=4, this indicates an inclination of
$i$$<$30$^\circ$, consistent with observations at all wavelengths
within a (limited) unified scheme. Assuming an average rotational
velocity for high-$z$ quasars, this even sets a limit on the mass of
the stellar bulge within the central 550\,pc of $M_{\rm
  bulge}$=3.4\,$M_{\rm BH}$, which falls by almost two orders of
magnitude short of the $M_{\rm BH}$--$M_{\rm bulge}$ relation observed
for nearby galaxies (see also discussion by Wei\ss\ \etal\
\citeyear{wei07}). We thus conclude that the SMBH in this distant
quasar is already in place, while the buildup of the stellar bulge is
still in progress. A similar indication was found for the $z$=6.4
quasar SDSS\,J1148+5251 (Walter \etal\ \citeyear{wal04}), the $z$=4.1
quasar PSS\,J2322+1944 (Riechers et al.\ \citeyear{rie08a}), and the
$z$=4.4 quasar BRI\,1335-0417 (Riechers et al.\ \citeyear{rie08b}),
and thus suggests a breakdown of this relation at early cosmic times
and/or toward the high mass end.

The observations presented herein highlight the importance of
spatially and dynamically resolved studies of molecular gas in
galaxies at $z$$\gtrsim$4 for our general picture of galaxy formation
and evolution. In the next decade, the Expanded Very Large Array
(EVLA) will enable us to commence such studies at improved
sensitivities and spectral resolution. The low $J$ transition studies
of molecular gas with the EVLA will be complemented by observations of
the higher $J$ transitions with the Atacama Large (sub-)Millimeter
Array (ALMA) at even higher spatial resolution. Both instruments will
allow us to probe significantly deeper down the luminosity function(s)
of the different populations of high redshift galaxies using molecular
gas studies, and thus largely improve our picture of galaxies in the
early universe.

\acknowledgments

We thank Dennis Downes, Axel Wei\ss, Rob Ivison, Melanie Krips, and
Vernesa Smol{\v c}i\'c for helpful discussions, and the anonymous
referee for a critical reading of this manuscript. We also thank Rob
Ivison for making his archival VLA data available, and Eiichi Egami
and George Chartas for providing near-infrared and X-ray images of
APM\,08279+5255.  DR acknowledges support from from NASA through
Hubble Fellowship grant HST-HF-01212.01-A awarded by the Space
Telescope Science Institute, which is operated by the Association of
Universities for Research in Astronomy, Inc., for NASA, under contract
NAS 5-26555, and from the Deutsche Forschungsgemeinschaft (DFG)
Priority Program 1177. CC acknowledges support from the
Max-Planck-Gesellschaft and the Alexander von Humboldt-Stiftung
through the Max-Planck-Forschungspreis 2005. The National Radio
Astronomy Observatory (NRAO) is operated by Associated Universities
Inc., under cooperative agreement with the National Science
Foundation. This research has made use of the NRAO Data Archive
System. Some of the data presented in this paper were obtained from
the Multimission Archive at the Space Telescope Science Institute
(MAST). Support for MAST for non-HST data is provided by the NASA
Office of Space Science via grant NAG5-7584 and by other grants and
contracts. This research has made use of NASA's Astrophysics Data
System.

\end{document}